\documentclass[aps,prm,reprint,superscriptaddress]{revtex4-1}

\usepackage{mhchem}
\usepackage{graphicx}
\usepackage{longtable}

\begin{document}


\title{Reversal magnetization, spin reorientation and exchange bias in \ce{YCrO_3} doped with praseodymium}


\author{A. Dur\'an}
\email[Corresponding author: A. Dur\'an \\ Phone: ++52 (646) 175 06 50 \\ Fax: ++52 (646) 175 06 51 \\ e-mail: ]{dural@cnyn.unam.mx}
\affiliation{Universidad Nacional Aut\'onoma de M\'exico, Centro de Nanociencias y Nanotecnolog\'ia, Km. 107 Carretera Tijuana-Ensenada. Apartado Postal 14, C. P. 22800, Ensenada, B. C. M\'exico.}
\author{R. Escamilla}
\author{R. Escudero}
\author{F. Morales}
\affiliation{Universidad Nacional Aut\'onoma de M\'exico, Instituto de Investigaciones en Materiales, Apartado Postal 70-360, Ciudad de M\'exico 04510, M\'exico. }
\author{E. Verd\'in}
\affiliation{Universidad de Sonora, Departamento de F\'isica, Apartado Postal 1626, Hermosillo, Sonora, M\'exico}

\date{\today}

\begin{abstract}
Crystal structure, thermal and magnetic properties were systematically studied in the \ce{Y_{1-x}Pr_xCrO_3} with $0 \leq x \leq 0.3$ compositions. Magnetic susceptibility and specific heat measurements show an increase of the antiferromagnetic transition temperature ($T_N$) as Pr is substituted in the Y sites and notable magnetic features are observed below $T_N$. Strong coupling between magnetic and crystalline parameters is observed in a small range of Pr compositions. A small perturbation in the lattice parameters by Pr ion is sufficient to induce a spin reorientation transition followed by magnetization reversal, to finally induce exchange bias effect. The spin reorientation temperature ($T_{SR}$) is increased from 35 K to 149 K for $0.025 \leq x  \leq 0.1$ compositions. It is found that the Cr spins sublattice rotates continuously from $T_{SR}$ to a new spin configuration a lower temperature. In addition, magnetization reversal is observed at $T^* \sim 35$ K for x= 0.05 up to $T^* \sim 63$ K for x = 0.20 composition.  The $M-H$ curves show negative exchange bias effect induced by Pr ions, which are observed below of 100 K and being more intense at 5 K. At 10 K, the magnetic contribution of the specific heat, as well as the ZFC magnetization, show the rise of a peak with increasing Pr content. The magnetic anomaly could be associated with the freezing of the Pr magnetic moment randomly distributed at the 4c crystallographic site. A clear correspondence between spin reorientation, magnetization reversal and exchange bias anisotropy with the tilting and octahedral distortion is also discussed.
\end{abstract}


\maketitle

\section{Introduction}
Complex oxides of transition metals with perovskite structure represent a fascinating playground for basic solid state research: new electronics and exotic ground states emerge via the competing interplay between spin, orbital, charge and lattice degrees of freedom \cite{salamon01}. High $T_c$ superconductivity, colossal magnetoresistance, coexistence and competition of magnetism and ferroelectricity are perhaps the most known examples. Recently, there is a renewed interest in rare earth manganites, ferrites, orthoferrites and orthochromites due to the coexistence or coupling between lattice and magnetic order parameters leading to magnetoelectric effect and striking/exotic magnetic properties \cite{kimura03,cheong07,eerenstein06,rajeswaran12,kumar15}. In these compounds, from the basic point of view, a growing interest has been focused on the microscopic interactions responsible for the interplay between the lattice and spin ground states \cite{bellaiche12}. This phenomenon is associated with a technological appealing potential for applications, as magneto-optic, spintronic and data storage devices \cite{ramesh07,bibes08,ramesh10}. From a general point of view, the orthochromites with formula \ce{RCrO_3} where R = Y or rare earth are iso-structural orthorhombic perovskite-derived structures (space group {\sl Pbnm}). The R substitution from La to Lu decreases the tolerance factor causing the cooperative octahedral rotation and consequently, the CrO6 octahedral tilting is progressively reduced. Accordingly, the antiferromagnetic order (AFM) temperature of the Cr$^{+3}$ sub-lattice is decreased from 282 K for \ce{LaCrO_3} to 112 K for \ce{LuCrO_3} \cite{prado13}. Below the  N\'eel temperature $T_N$, these compounds present a weak ferromagnetism (WFM) arising from a slight canting of the AFM spins that lie either along the a-axis or c-axis of the unit cell \cite{yamaguchi73}. The WFM results from an antisymmetric superexchange between Cr$^{+3}$ spins, also known as Dzyaloshinsky-Moriya (D-M) interaction \cite{moriya,treves}. The D-M exchange interaction is the main responsible for coupling of the spin and the lattice degrees of freedom \cite{cheong07}. Interestingly, when the crystal contains both $d-$ion subsystem and $f-$ion subsystem, the D-M interaction breaks the inversion symmetry through the incommensurate magnetization on the $d-$ion subsystem at $T_N$ \cite{sergienko06}. On the other hand, the magnetic behavior becomes more complex as the magnetization of the $f-$ion subsystem increases below $T_N$. Accordingly, a rich variety of magnetic and electric properties, such as magnetostriction induced polarization, spin reorientation, magnetization reversal and exchange bias is reported in several orthorhombic manganites, orthochromites and orthoferrites \cite{cheong07,kumar15,sergienko06}. Different combinations of rare earth and transition metal ions including a relative concentration between them can produce ferroelectric polarization at the magnetic ordering and a characteristic behavior called magnetization reversal (MR). MR means that the magnetization turns to diamagnetic state a certain temperature (compensation temperature, $T^*$) under low applied magnetic field \cite{yoshii00,khomchenko08,mao11}. This phenomenon is not exclusive of these compounds but rather of a larger group of materials such as spinels, garnets, orthovanadates and Prussian blue analogs \cite{gorter53,pauthenet58,ren98,ohkoshi97}. Magneto-reversal behavior can be explained using the N\'eel criterion; however, for others such as \ce{YVO_3}, the competition between D-M interaction and a single-ion magnetic anisotropy (SIMA) has been suggested as the most viable explanation \cite{ren00,ohkoshi99}. The MR in orthochromites is a clear manifestation that the magnetic ground states are instable and susceptible to small perturbations caused by the strong competition between $f-$ion and $d-$ion subsystems and by octahedral tilting arising from the R partial substitution. This last aspect has been little addressed in recent research on these phenomena.

In this report, we provide a detailed study of the structural, thermal and magnetic properties of the \ce{Y_{1-x}Pr_xCrO_3} solid solution. The structural and magnetic measurements indicate that the Pr substitution has a notable effect on the magnetic ground state. Spin reorientation (SR) and reversal magnetization are very sensitive to Pr substitution in a narrow range of Pr composition. These facts not only suggests an energetic condition for the presence of SR and MR but also the development of ferromagnetic domains coexisting with antiferromagnetic domains, which produce the development of exchange magnetic anisotropy also called exchange bias (EB). The close connection between octahedral distortion and these complex magnetics ground state are analyzed and discussed.

\section{Experimental Details}

Self-propagating high-temperature synthesis (combustion) method was used to produce polycrystalline \ce{Y_{1-x}Pr_xCrO_3} with $0.0 \leq x \leq 0.3$ solid solution samples. Stoichiometric amounts of precursor nitrates \ce{Y(NO_3)_{3} \cdot 6H_2O} (99.8\% Alfa-Aesar), \ce{Pr(NO_3)_3 \cdot 6H_2O} (99.9\% Alfa-Aesar) and \ce{Cr(NO_3)_3 \cdot 9H_2O} (99.9\% Sigma-Aldrich) were dissolved in 2-methoxyethanol as a fuel and distilled water to form the precursor solution. The synthesis details have been reported in elsewhere \cite{duran10}. The fine as-combustion powders were then ground and thermally treated in a furnace in a single step process: a heating rate of 10$^\circ$ C min$^{-1}$ to reach 1200$^\circ$ C for 6 h. Phase identification of the samples was done with an X-ray Siemens D5000 diffractometer using Co-K$_\alpha$ radiation and a Fe filter. Intensities were measured at room temperature in steps of 0.02$^\circ$, for 14 seconds, in the $2\theta$ range 10$^\circ$ - 100$^\circ$. The crystallographic phases were identified by comparison with the X-ray patterns of the JCPDS database. The crystallographic parameters were determined using a Rietveld refinement program, MAUD v 1.7.7 with multi-phase capability \cite{maud}. The specific heat measurements were carried out on PPMS (Physical Property Measurement System, Quantum Design) at temperatures from 2 to 300 K. Magnetization was measured with a SQUID- based magnetometer (MPMS-5T by Quantum Design). The susceptibility measurements were performed at 1 kOe in the zero field cooled (ZFC) and field cooled (FC) modes (2-300 K) while the magnetization vs applied magnetic field measurements was performed at   $\pm5$ Tesla at 5, 50 and 100 K.

\begin{figure}[t]
\begin{center}
\includegraphics[scale=0.4]{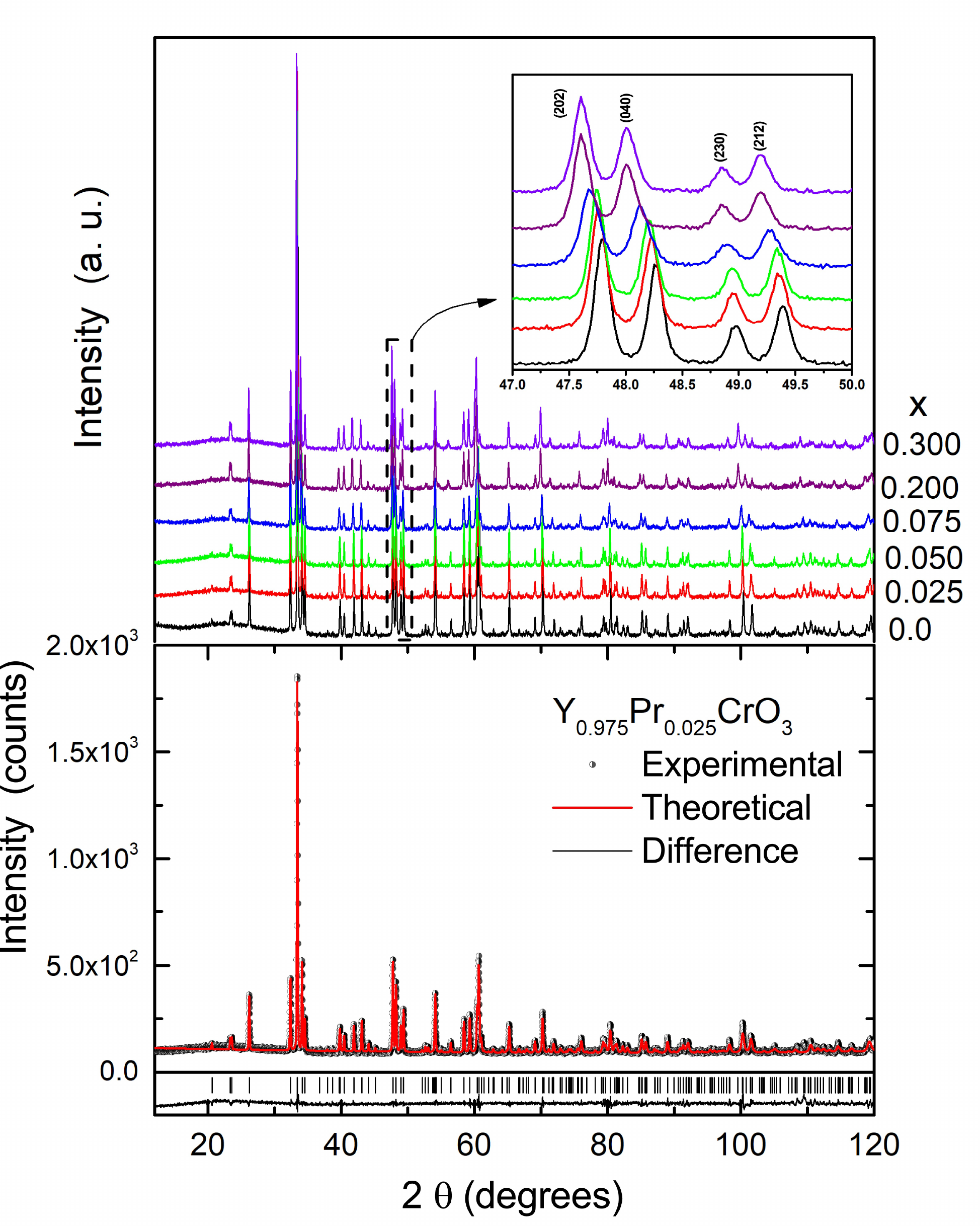}
\caption{Upper panel: X-ray diffraction patterns for the \ce{Y_{1-x}Pr_xCrO_3} with $0 \leq x \leq 0.3$ compositions. The inset shows the shift of the plane (202), (040), (230) and (212) as Pr content. Bottom panel: The fitting results of the Rietveld analysis for the x=0.025 sample along with experimental ($\bullet$), calculated ($-$) and the bottom line is the difference between the observed and calculated patterns.}
\end{center}
\end{figure}

\begin{figure}[t]
\begin{center}
\includegraphics[scale=0.38]{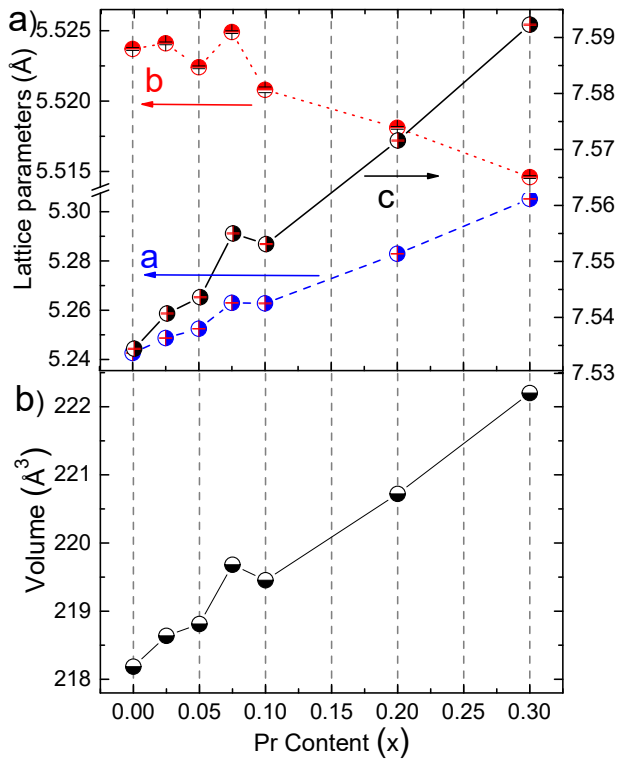}
\caption{(a) Lattice parameters are in \AA in both y-axes and (b) shows the unit cell volume of the orthorhombic structure as a function of Pr content.}
\end{center}
\end{figure}

\begin{figure}[h]
\begin{center}
\includegraphics[scale=0.4]{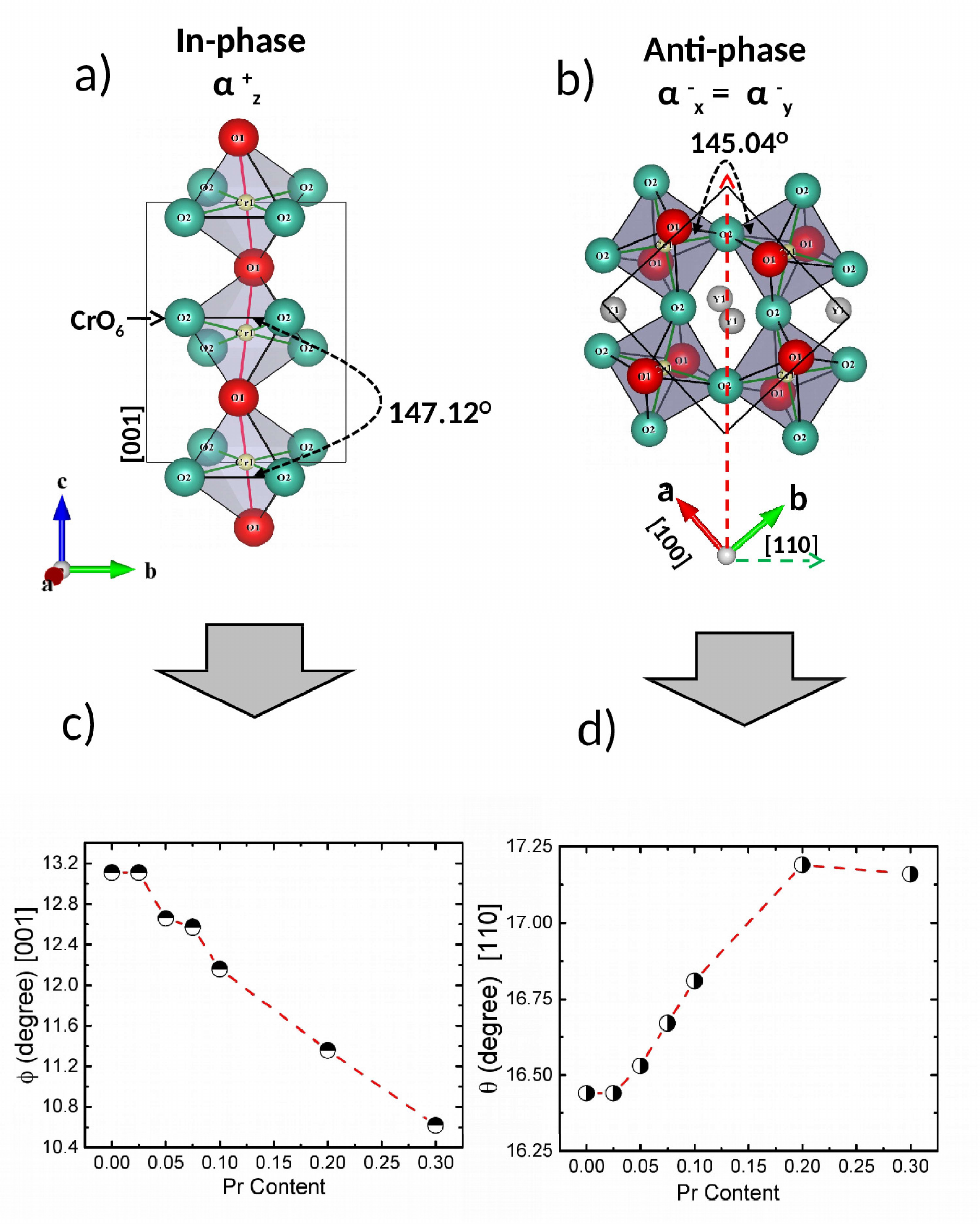}
\caption{The octahedral arrangement of the orthorhombic $Pbnm$ ($\alpha^- \alpha^- \alpha^+$) phase of \ce{Y_{1-x}Pr_xCrO_3} perovskite. a) In-phase $\alpha^+$ and b) anti-phase $\alpha^- = \alpha^-$ tilting about [001] and [110] axes of the \ce{CrO_6} polyhedron, respectively. The tilting angles along the [001] and the [110] directions also is displayed in c) and d) graph.}
\end{center}
\end{figure}

\section{RESULTS}

\subsection{Structural analysis}

Fig. 1 shows the X-ray powder diffraction patterns for \ce{Y_{1-x}Pr_xCrO_3} with $0.0 \leq x \leq 0.3$ compositions. The analysis of the data shows a single phase for all samples corresponding to the distorted perovskite structure with orthorhombic symmetry \ce{YCrO_3} (ICSD n$^\circ$ 34-0365). The X-ray diffraction patterns of the samples were Rietveld-fitted using a space group $Pbnm$ (No. 62), considering the possibility that Pr occupies Y sites. As an example, the profile fitting of the x-ray diffraction pattern for the \ce{Y_{0.975}Pr_{0.025}CrO_3} sample is shown in the bottom panel of Fig.1. The crystallographic parameters obtained from the Rietveld refinements are shown in Table I. From the refinement results, the lattice parameters and the unit cell volume behavior with Pr content are seen in Fig. 2. For the undoped sample, the lattice parameters values are in agreement with other published results \cite{prado13,duran10}. The $a-$ and $c-$axes show a significant increase with increasing Pr content, while the $b-$axis shows a slight decrease. The net result is an increase in cell volume with increasing Pr content, which is due to the effective ionic radius of Pr$^{+3}$ (1.126 \AA) being larger than that of the Y$^{+3}$ (1.019 \AA) ion with eight-coordination \cite{shannon76}. Note that the lattice parameters and the volume reveal a small anomaly at about x=0.075 of Pr content. This behavior appears to be inconsistent with the criterion of chemical pressure effect. Thus, to get a more precise understanding of this behavior, the internal crystallographic parameters (octahedral distortion, bond-length, bond-angle and tilting) were extracted from the Rietveld refinement and are listed in Table II. The internal structural parameters such as the octahedral distortion ($\Delta$) and the $<Cr-O-Cr>$ bond angles are affected by the Y/Pr substitution whereas the average $<Cr-O>$ bond lengths are rigid remaining almost constant. This fact is an expected result for Cr$^{+3}$ in octahedral environment \cite{arevalo09,duran12}. The Cr-O-Cr angles along the [001] and [110] direction correspond to the in-phase octahedral tiltings $\alpha_z^+$ and anti-phase octahedral tilting $\alpha_x^- = \alpha_y^-$  respectively, as is seen in Fig. 3 a, b). There, the apical oxygen atoms are denoted as O(1) and the equatorial oxygen atoms as O(2) in the \ce{CrO_6} octahedral perovskite. Both octahedral tiltings can be calculated using the expressions $\theta  = (180-<Cr-O(1)- Cr>)/2$ and $cos \phi   = cos((180-<Cr-O(2)-Cr>)/2) /\sqrt{cos \theta})$ \cite{zhao93}. Here, the tilt angles $\phi$[001] and $\theta$[110] for x=0 are in agreement with the values reported in ref.\cite{prado13,sardar11}. Fig 3 c and d) show an increase in the tilt $\theta$  angle and a decrease in the tilt $\phi$  angle with increasing the Pr content. Concomitantly, the continuous deviation of the in-phase and anti-phase tilting angles and the octahedral distortion with Y/Pr substitution should produce a strong influence on the magnetic properties as will be discussed in what follows.

\begin{table*}
\caption{Structural parameters and atomic positions for \ce{(Y_{1-x}Pr_x)CrO_3} system at room temperature. }
\begin{ruledtabular}
\begin{tabular}{lcllllllll}
 &x=       &   & 0.00 & 0.025 & 0.050 & 0.075  & 0.100  & 0.200  & 0.300   \\ \hline
 & $a$(\AA) &   & 5.2426(1) &5.2487 (1) &5.2525(1)&5.2630 (1)&	5.2628 (2)&	5.2829(1)&	5.3052(1) \\
 & $b$(\AA)&     &5.5237(1)& 5.5241(1)&	5.5224(1)&	5.5249 (1)&	5.5208 (2)&	5.5180(1)&	5.5146(1) \\
 & $c$(\AA)&	 &7.5344(1)&7.5407 (1)&	7.5436(1)&	7.5550 (2)&	7.5531 (3)&	7.5716(1)&	7.5923(1) \\
 & V(\AA$^3$)&	 &218.19&	218.64&	218.81&	219.68&	219.45&	220.72&	222.12   \\  \hline
Y& &x	&-0.0169(4)&	-0.0172(3)&	-0.0176(3)&	-0.0178(2)&	-0.0182(3)&	-0.0195(3)&	-0.0208(3) \\
 &	&y	&0.0664(2)&	0.0655(2)	&0.0647(1)&	0.0639(1)	&0.0630(2)&	0.0596(2)&	0.0562(2)   \\
 &	& B(\AA$^2$)	&0.13(3)&	0.17(4)&	0.10(2)&	0.17(3)	&0.31(4)	&0.39(5)	&0.37(2) \\
Cr&	&B(\AA$^2$)	&0.11(4)	&0.14(3)	&0.10(3)	&0.16(5)	&0.26(3)	&0.26(3)	&0.29(1)  \\
O(1)&	&x	&0.099(2)&	0.099(1)&	0.100(2)&	0.101(1)&	0.102(2)&	0.105(2)&	0.105(2)  \\
	&  &y	&0.464 (2)	&0.464(1)	&0.465 (1)	&0.465(2)	&0.465(1)	&0.466(1)	&0.467(1)  \\
	&  &B(\AA$^2$)	&0.25(3)	&0.39(1)	&0.25(2)	&0.25(2)	&0.46(3)&0.20(3)	&0.25(3)  \\
O(2)&	&x	&-0.306 (1)	&-0.306(1)	&-0.305(1)	&-0.305(1)	&-0.304(1)	&-0.303(2)	&-0.301(2)  \\
	&   &y	 &0.306(1)	& 0.306(1)	& 0.305(1)	 &0.305(2)	 &0.305(1)	&0.303(1)	&0.302(1)  \\
	&   &z	 &0.056(1)	& 0.056(1)	 &0.055(1)	 &0.055(1)	 &0.054(1)	 &0.053(1)	&0.051(1)  \\
	&   &B(\AA$^2$)	&0.27(2)	&0.37(2)	&0.20(1)	&0.19(2)	&0.25 (3)&0.29(5)&0.19(2)  \\ \hline
& &$R_b$ (\%)    & 3.9 & 3.8 & 4.3 & 4.0 & 3.7 & 3.6 & 3.6 \\
& &$R_{wp}$(\%)  & 5.1 & 5.1 & 5.7 & 5.1 & 4.9 & 4.6 & 4.5 \\
& &$R_{exp}$(\%) & 2.8 & 3.0 & 3.1 & 3.1 & 3.0 & 2.9 & 2.9 \\
& &$\chi^2$(\%)	 & 1.8 & 1.7 & 1.8 & 1.6 & 1.6 & 1.6 & 1.6 \\
\end{tabular}
\end{ruledtabular}
Space group: $Pbnm$.  Atomic positions: Y: 4c (x, y, 0.25); Cr: 4b (0, 0.5, 0); O(1): 4c (x, y, 0.25) and O(2):  8d (x, y, z).
\end{table*}

\begin{table*}
\caption{Geometrical parameters characterizing the crystal structure of (\ce{Y_{1-x}Pr_x)CrO_3} system. The octahedral distortion parameter $\Delta$  of a coordination polyhedron \ce{BO_N} with an average bond length B-O $<d>$, is defined as $\Delta=(1/N)\sum_{n=1,N} {(d_n-<d>)/<d>}^2$ \cite{alonso00}. The tilt angles $\phi$ and $\theta$  of \ce{CrO_6} octahedral around pseudocubic [001] and [110] direction are  obtained from the two angles; $\theta_1$ and $\theta_2$ \cite{zhao93}.}
\begin{ruledtabular}
\begin{tabular}{llllllll}
x=       & 0.000	&0.025	   & 0.050	  &0.075	 & 0.100    &0.200	   &0.300     \\  \hline
Cr-O1:2  & 1.964(3) & 1.966(2) & 1.967(3) & 1.972(2) & 1.971(2) & 1.976(3) & 1.973(2) \\
Cr-O2:2  & 1.975(3) & 1.977(3) & 1.974(2) & 1.977(3) & 1.973(3) & 1.981(3) & 1.987(3) \\
Cr-O2:2  & 2.017(2)	& 2.018(3) & 2.015(3) & 2.016(3) & 2.016(3)	& 2.010(2) & 2.010(2) \\
<Cr-O2>  & 1.996	& 1.997	   & 1.995	  & 1.996	 & 1.995	& 1.996	   & 1.999    \\
<Cr-O> 	 & 1.985	& 1.987	   & 1.985	  & 1.988	 & 1.987	& 1.989	   & 1.990    \\
$\Delta$(Cr-O)$\times10^{-4}$ & 2.64	& 2.53	& 2.27	& 1.96	& 2.18	& 1.13	& 1.17    \\
$\theta_1$:Cr-O(1)-Cr &147.12(2)&147.11(2)&146.94(3)&146.66(2)&146.37(3)&145.62(3)& 145.69(3)  \\
$\theta_2$:Cr-O(2)-Cr &145.04(2)&145.03(3)&145.62(2)&145.61(3)&146.05(3)&146.77(3)&147.79(3) \\
$\phi$[001]&	13.11&	13.11&	12.66&	12.57&	12.16&	11.36&	10.62  \\
$\theta$[110]&	16.44&	16.44&	16.53&	16.67&	16.81&	17.19&	17.16  \\
\end{tabular}
\end{ruledtabular}
\end{table*}

\subsection{Magnetic results}

\subsubsection{Antiferromagnetic regime}

Temperature-dependence of the ZFC and FC magnetic susceptibility under an applied field of 1 kOe for all samples ($0 \leq x \leq 0.3$) is shown in Fig. 4. Note that the Y/Pr substitution gives rise to a striking and unusual development in the magnetic ground states. For x=0, the susceptibility shows a large splitting in the ZFC and FC curves below $\sim 142$ K (see inset of Fig. 4 a). It has been reported that the antiferromagnetic G-type structure with spin canting of the Cr$^{+3}$ (S=3/2) ion occurs below $\sim 142$ K, for \ce{YCrO_3} \cite{judin66}. The substitution of Pr at the Y site produces an increase of the $T_N$ up to 166 K for x = 0.30. Moreover, the magnetic susceptibility curves show intriguing magnetic properties such as spin reorientation (SR) and temperature induced reversal magnetization (MR) at $T^* < T_{SR} <  T_N$. The susceptibility curves show a splitting of ZFC and FC mode followed by a sudden drop of the magnetization at $\sim 35$ and $\sim 105$ K for x=0.025 and 0.050, respectively. This is a signature of spin reorientation transition occurring at $T_{SR}$. The hysteresis in the FC and ZFC magnetization data below SR suggests a first-order transition (arrows in Fig. 4 b) in a similar way as was observed in the \ce{YFe_{1-x}Mn_xO_3} compound \cite{mandal13}. For higher Pr content, the TSR is shifted to a higher temperature from $\sim 35$ K for x=0.025 to $\sim 145$ K for the x=0.10 composition. Below $T_{SR}$, a characteristic point in the FC curve is the negative magnetization or the induced MR, which occurs to at relatively high applied magnetic field (1 kOe) in the $0.050 \leq x \leq 0.20$ composition range. Cooling in an applied field of 1 kOe, the magnetization shows a maximum with a positive magnetization, and magnetization reversal is observed at compensation temperatures ($T^*$) of 17, 31, 45 and 63 K for x=0.05, 0.075, 0.10 and 0.20 compositions to finally vanish for x=0.30 of Pr content (Figs. 4 b-d). The magnetization takes a value of about -0.5 emu/mol for x=0.2 Pr content at 2 K. The negative magnetization value is similar to that obtained in the equimolar \ce{La_{0.5}Pr_{0.5}CrO_3}, but in a weaker applied magnetic field (100 Oe) \cite{yoshii00}. It is also noted that the magnetization reversal in \ce{Y_{1-x}Pr_xCrO_3} occurs in narrower range of Pr content, between $0.050 \leq x \leq 0.2$, while the reversal magnetization in \ce{La_{1-x}Pr_xCrO_3} occurs in a broader composition range of composition ($0.2 \leq x \leq 0.8$) \cite{yoshii01}. At this point, a question that must be addressed is why both SR and MR phenomena occur in a narrow range of Pr composition ($0.025 \leq x \leq 0.20$). It is worth noting that in this narrow range of Pr compositions occurs an anomalous octahedral distortion in close connection with the $t-e$ hybridization and magnetization behavior. These facts not only suggests energy condition for the presence of both phenomena but also the development of ferromagnetic domains coexisting with antiferromagnetic domains which favor the development of the exchange bias effect at lower temperatures as is indicated by the $M-H$ curves in the following section.

\begin{figure}[h]
\begin{center}
\includegraphics[scale=0.4]{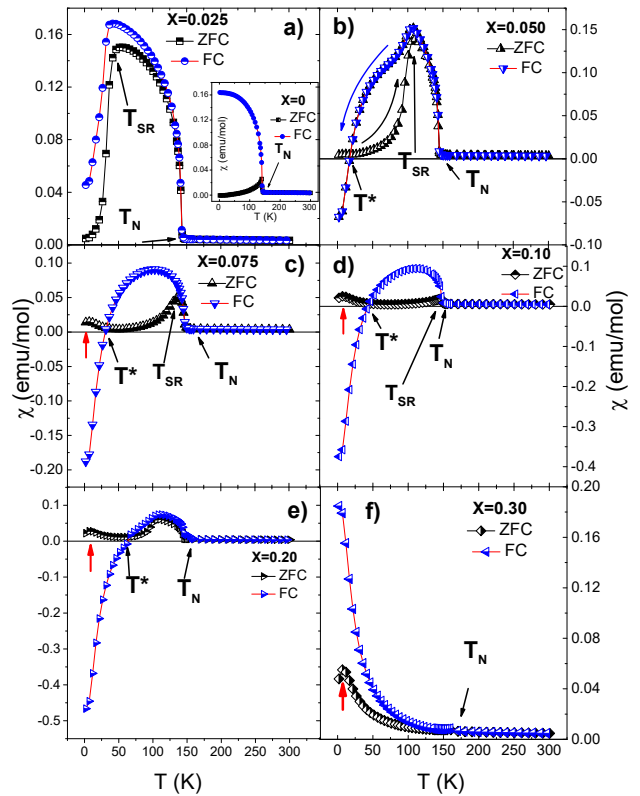}
\caption{Magnetic susceptibility, ZFC and FC cycles, at 1 kOe from a) x=0.025 to f) x=0.30 compositions. In the panel graphs, the $T_N$, $T_{SR}$ and $T^*$ are the AFM transition, spin reorientation and compensation temperatures, respectively. Inset of panel a) shows the magnetic susceptibility for of a the pristine sample. The continuous line is a guide to the eye.}
\end{center}
\end{figure}

\begin{figure}[h]
\begin{center}
\includegraphics[scale=0.4]{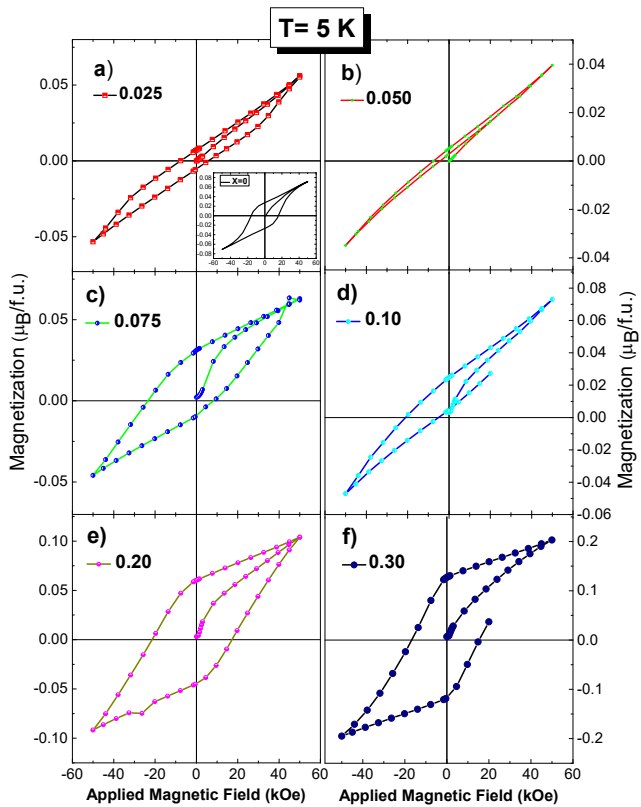}
\caption{$M-H$ hysteresis loops measured at 5 K for \ce{Y_{1-x}Pr_xCrO_3} with $0\leq x \leq 0.3$ solid solution. The inset of the upper panel shows to hysteresis loop for the pristine sample.}
\end{center}
\end{figure}

\begin{figure}[h]
\begin{center}
\includegraphics[scale=0.4]{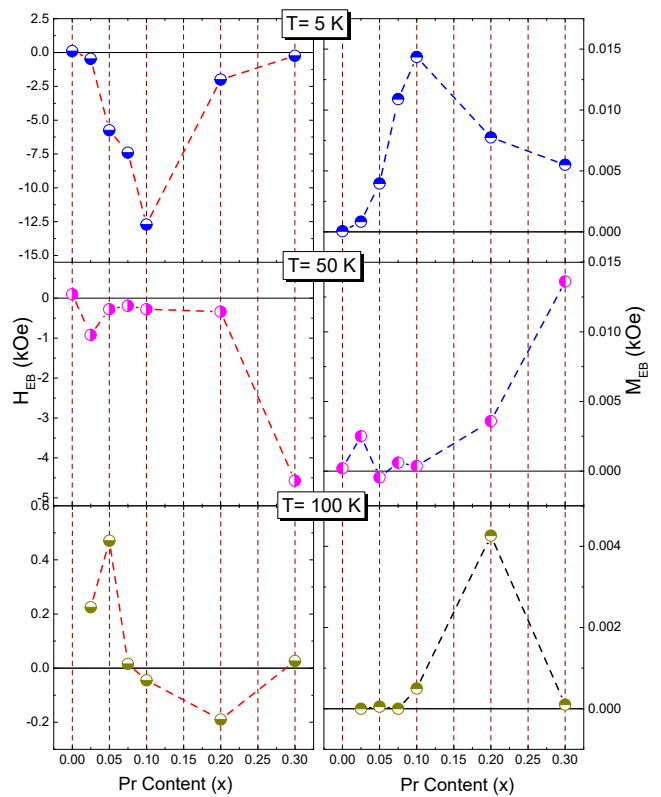}
\caption{$H_{EB}$ and $M_{EB}$ as a function of the Pr content at 5, 50 and 100 K.}
\end{center}
\end{figure}

\subsubsection{M-H hysteresis loops.}

Magnetization as a function of the applied magnetic field ($\pm$50 kOe) was measured at 5 K, after cooling from room temperature (through $T_N$) without an applied magnetic field for all compositions. Fig. 5 shows the hysteresis loops for each composition measured at 5 K. As expected, the pristine sample shows a hysteresis loop due to spin canting AFM ordering with the coercive field ($H_c$) and the remanent magnetization ($M_r$) of $\pm$16 kOe and 0.027 $\mu_B$/f.u. at 5 K , respectively (see inset Fig. 5 a) in agreement with the values reported in ref. \cite{duran10,serrao05}. A drastic decrease of both, the $H_c$ and $M_r$ for x=0.025 and 0.050 of Pr content is observed (Fig 5 a, b). After that, the coercive field ($H_c$) increases and then decreases for x=0.075 and 0.10 compositions, respectively. Finally, for the x=0.20 and 0.30 compositions, the $H_c$ and $M_r$ take values higher than the pristine sample. The result shows a strong ferromagnetic contribution for these last two compositions at 5 K. On the other hand, the $M-H$ curves also show a shift of the hysteresis loop towards negative applied fields axis from x=0.025 to x=0.20 composition. This fact indicates exchange bias (EB) induced by Pr substitution. It is well known that the EB effects arise when the FM and AFM domains are coupled through an interface \cite{kumar15,nogues99}. The results here suggest that both FM and AFM domains in close contact lead to an additional anisotropy (showed by asymmetric, $H_c$) via exchange coupling in the bulk-doped samples.  The negative shifting in the hysteresis curves are clearly seen from x=0.050 to 0.10 Pr content at 5 K. For higher compositions, the EB tends to vanish. The behavior of the exchange bias field $H_{EB}$, and the remanent magnetization $M_{EB}$, as a function of Pr content for 5, 50 and 100 K are plotted in Fig. 6. These values were determined for each concentration using the relation $H_{EB} = (H_+ + H_-)/2$ and $M_{EB} = (M_+ + M_-)/2$ taking the H and M values during ascending and descending branches of the hysteresis loops. At 5 K, the negative $H_{EB}$ is near zero for x=0.05 and continuously decreases down to ~0.13 kOe at x=0.10; after that composition, the $H_{EB}$ increases to almost vanish at x=0.30 Pr content. In a similar way, the $M_{EB}$ continuously increases with a maximum at x = 0.10 and then decreases for x=0.30 of Pr content as seen in the upper right panel of Fig. 6.  Similarly, the $M_{EB}$ and $_{HE}$B persist, but they are small at 50 K. $H_{EB}$ and the $M_{EB}$ are slightly increased for x=0.2 and 0.3. Finally, a turn from negative to positive $H_{EB}$ is observed at about x= 0.1 at 100 K as is shown in the lower panel of Fig 6.  It is worth noting that the $H_{EB}$ and the $M_{EB}$ display strong changes below $T_{SR}$ and $T^*$ with maximum values at x=0.10 at 5 K.  The competition of the FM and AFM domains, presumably due to Pr and the Cr interaction are clearly appreciated at 5 K where Pr-O-Cr interactions become stronger. A maximum in the $-H_{EB}$ and $M_{EB}$ indicates that the canted AFM moments align with stronger intensity opposite to the applied field giving rise to a negative exchange bias (NEB) \cite{roshchin05} at x=0.1.

\begin{figure}[h]
\begin{center}
\includegraphics[scale=0.4]{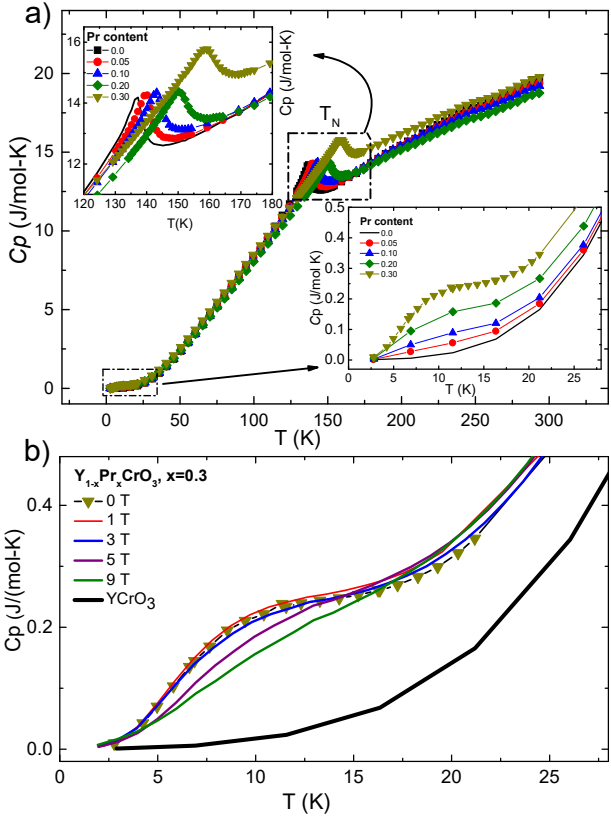}
\caption{a) Temperature dependence of the specific heat $C_P$ for the \ce{Y_{1-x}Pr_xCrO_3} solid solution. The upper and lower insets shows an expanded scale around $T_N$ and the development of an anomaly at about 10 K. b) Specific heat for the x=0.3 sample under applied magnetic field up to 9 T. For comparison the pristine sample is shown (thick line).}
\end{center}
\end{figure}

\subsubsection{Specific heat measurements.}

Fig. 7 a) shows the temperature dependence of total specific heat, $C_P$ for \ce{Y_{1-x}Pr_xCrO_3} with x=0.0, 0.05, 0.1, 0.2 and 0.3 from 300 down to 2 K. Two anomalies are observed in $C_P$, which are clearly observed in the expanded scale of the upper and lower inset of Fig. 7 a). The first anomaly occurs at $T_N$, which is indicative of the second-order AFM transition of the Cr ($d^3$, S=3/2) magnetic moments. The upper inset shows an increase of AFM transition from 142 to 166 K in agreement with the magnetic measurement. The second anomaly is displayed at low temperature as is seen in the lower inset of Fig. 7 a). The anomaly observed at about 10 K begins to increase with as the Pr content to finish as a hump for x=0.3. This feature resembles a Schottky anomaly. Fig. 7 b) shows the $C_P$ vs $T$ curves from 30 down to 2 K for x = 0.3 sample measured up to 9 T.  There, the $C_P$ vs $T$ curve for the pristine sample is shown for comparison. Under applied magnetic field, the hump is smoothed until almost it disappears at 9 T. The results ruled out the Schottky-like transition \cite{bartolome00} induced by Pr substitution.  On the contrary, the results clearly support that the magnetic anomaly at 10 K is associated with the Pr ions in the \ce{YCrO3} matrix. To evaluate the magnetic contribution near $T_N$, the phonon contribution is estimated (and subtracted from the $C_P$) from the Debye formula,
\begin{equation}
C_{lat}=9RN\big(\frac{T}{\theta_D}\big)^3 \int_{0}^{\theta_D/T}\frac{x^4 e^x}{(e^{x}-1)^2} dx         \end{equation}

Where, $N = 5$ is the number of atoms per cell, R= 8.314 J/mol-K is the ideal gas constant, and $T_D$ is the Debye temperature. The Debye function reproduces the experimental data above 235 K with $T_D = 655$ K. This function is shown as a continuous line in Fig. 8 a). There, the magnetic contribution is also plotted as $Cm/T$ for \ce{YCrO3}. The Debye temperature ($T_D$) values obtained for the different Pr concentrations are between 645 K and 655 K. In Fig. 8 b) we plot the magnetic contribution, $Cm/T$ for \ce{Y_{1-x}Pr_xCrO_3} for x = 0, 0.05, 0.10, 0.20 and 0.30. The $Cm/T$ shows: i) the $T_N$ anomaly shifts towards higher temperatures as x increases in agreement with magnetic measurements, ii) a broad peak at about 60 K whose magnetic contribution increases slightly with increasing Pr content and iii) the arising of a peak at  10 K with increasing Pr content. Recently, Y. Sharma et al. \cite{sharma14} showed an additional phonon anomaly at 60 K likely related to the spin dynamic in the \ce{YCrO_3} compound. The large magnetic contribution was associated with metastable spin reorientation condition, which could be induced by applied magnetic field at $\sim 60$ K, as was reported by I. S. Jacobs et al. \cite{jacobs71}. However, this assumption can not be supported by our specific heat results since the $T_{SR}$ increases beyond 60 K with Pr content as seen in Fig. 4. This fact ruled out the connection of spin reorientation with the anomaly at 60 K in $Cm/T$. On the other hand, the evolution of the peak at about 10 K (Fig. 8 b), which only exists in \ce{Y_{1-x}Pr_xCrO_3} strongly suggest that it is related to the Pr-Pr exchange interaction, as was pointed out by T. Yamaguchi \cite{yamaguchi74}, which becomes visible at low temperature ($\sim 10$ K). However, this assumption is not valid for a diluted magnetic structure where the Pr ions are chemically disordered at the $4c$ crystallographic site.

\begin{figure}[t]
\begin{center}
\includegraphics[scale=0.4]{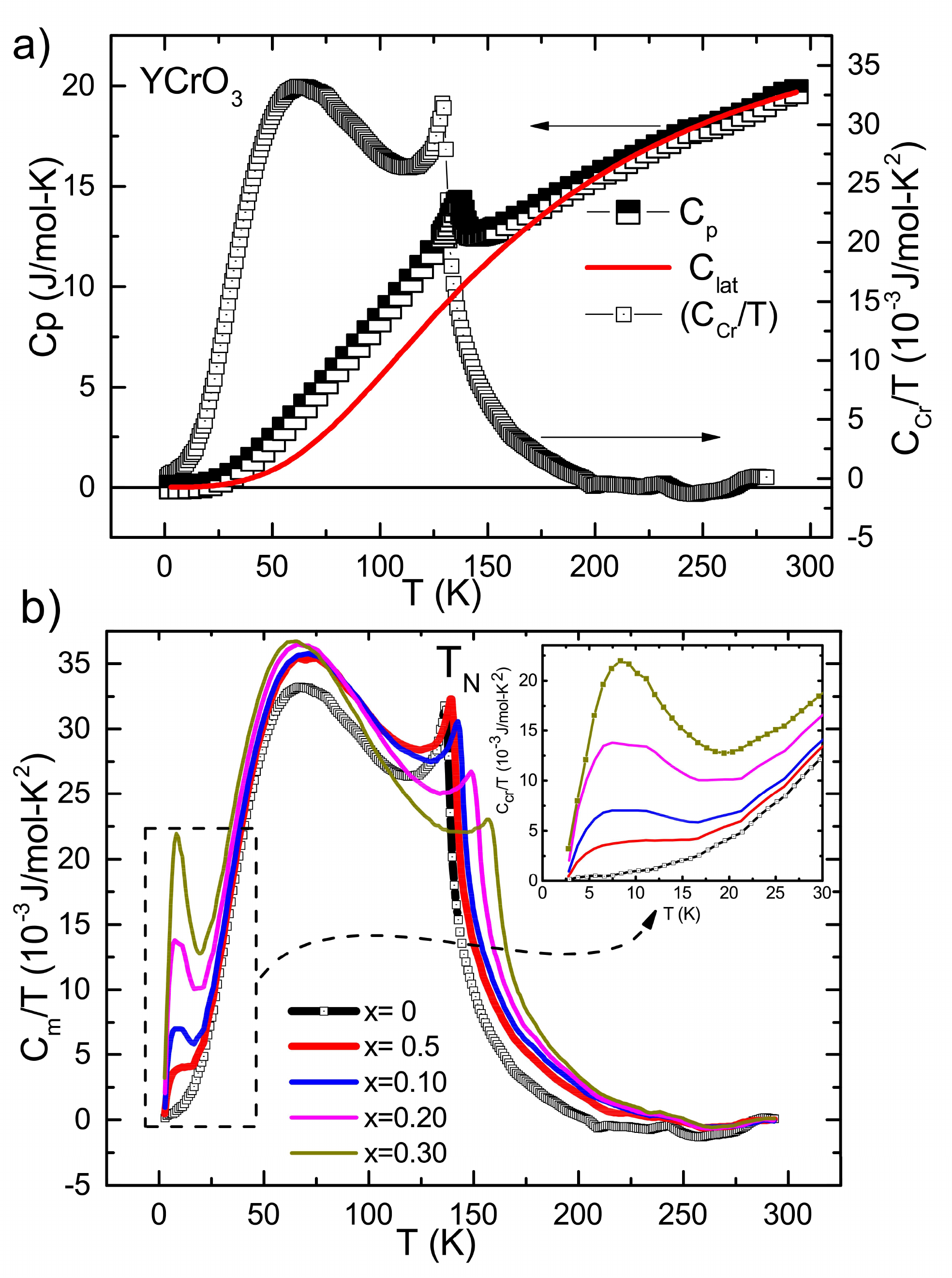}
\caption{a) The fitting result (solid line) and experimental data (half filled square) for \ce{YCrO_3} using the Debye specific heat function. The magnetic contribution $Cm/T$ is depicted with open squares. b) The $Cm/T$ is shown for x=0, 0.05, 0.1, 0.2 and 0.3 compositions. The inset shows the development of the hump at about 10 K.}
\end{center}
\end{figure}

\section{Discussion.}

Experimental results reveal several changes in the magnetic properties induced by the partial substitution of Y by Pr ions. The unfilled $f$-shell of Pr ions promotes the magnetic ground state toward more complex magnetic structures as a result of the anisotropic magnetic interactions of the Cr$^{+3}$ (S=3/2) and Pr$^{+3}$ (S=1) spins in the \ce{Y_{1-x}Pr_xCrO_3} ($0 \leq  x \leq  0.3$) solid solution. The magnetic moments of the Cr$^{+3}$ ions in the octahedral site are subjected to certain rotational forces arising through the exchange interaction between Cr$^{+3}$-Cr$^{+3}$, Cr$^{+3}$-Pr$^{+3}$ and, the weaker Pr$^{+3}$-Pr$^{+3}$ exchange interaction. Under these conditions, the magnetic structure is controlled by the crystal symmetry. The first complex magnetic structure is temperature-induced spin reorientation transition with Pr doping. In order to discern the possible magnetic structure during the spin reorientation, the G-type AFM configuration according to Bertaut notation \cite{bertaut68} as well as the experimental $M(H)$ curves are is taken into account. The \ce{RCrO_3} family allows $\Gamma_1$ (A$_x$, G$_y$, C$_z$), $\Gamma_2$ (F$_x$, C$_y$, G$_z$) and $\Gamma_4$ (G$_x$, A$_y$, F$_z$) ground state configurations. The $\Gamma_4$ (G$_x$, A$_y$, F$_z$) ground state remains weakly ferromagnetic below $T_N$ as it occurs with La and Y ions \cite{rajeswaran12}. When the R ion is magnetic \cite{gordon76,meltzer70} not only $\Gamma_4$ (G$_x$, A$_y$, F$_z$) spin configuration is possible but also $\Gamma_1$ (A$_x$, G$_y$, C$_z$) and $\Gamma_2$ (F$_x$, C$_y$, G$_z$) spin configurations. The $\Gamma_2$ (F$_x$, C$_y$, G$_z$) configuration also presents weak ferromagnetism such as occurs in \ce{TbCrO_3}, \ce{HoCrO_3}, and \ce{DyCrO_3} compounds \cite{yamaguchi74,shamir81}. Contrary to this magnetic configuration, the $\Gamma_1$ (A$_x$, G$_y$, C$_z$) configuration does not allow weak ferromagnetism as it was reported in \ce{ErCrO_3} where $\Gamma_4$ (G$_x$, A$_y$, F$_z$) changes to the non-magnetic $\Gamma_1$ (A$_x$, G$_y$, C$_z$) configuration below $T_{SR} \sim 22$ K \cite{grant69,kaneko77}. In the present study, the induced spin-reorientation transition is shifted to higher temperatures upon Pr substitution (Fig. 4 a, b) and, experimentally it is seen as a sudden drop in the magnetization followed by a splitting of ZFC and FC magnetic susceptibility. For example, the $T_{SR}$ increases from 35 to 105 K for x=0.025 and 0.050. The results also show that the magnetic hysteresis loop diminishes for x=0.025 concerning the pristine sample and almost disappears for x=0.050 at 5 K as is seen in Fig. 5 a-b). This magnetic behavior suggests that the Cr-spin rotates from weak ferromagnetic $\Gamma_4$ (G$_x$, A$_y$, F$_z$) to nonmagnetic $\Gamma_1$ (A$_x$, G$_y$, C$_z$) configuration below $T_{SR}$. For x=0.025, the reduced hysteresis loop suggests that the $\Gamma_1$ (A$_x$, G$_y$, C$_z$) configuration is incomplete at 5 K. For x=0.050, the spin reorientation transition temperature is increased to 105 K and an almost pure collinear AFM phase is obtained which is confirmed by the vanishing of the hysteresis loop at 5 K in Fig. 5 b); i.e., the $\Gamma_1$ (A$_x$, G$_y$, C$_z$) configuration is attained. The spin dynamic configuration as a function of temperature is more complex with higher Pr substitution, particularly at low temperatures since the Pr$^{+3}$-O-Cr$^{+3}$ magnetic interaction becomes important and may overcome the crystalline anisotropic forces.  To provide a better understanding of the spin dynamic configuration for the x=0.075 and 0.1 compositions, the inverse susceptibility data with the corresponding magnetic hysteresis loops at 5, 50 and 100 K are displayed in Fig. 9 a-b). For x=0.075, the splitting of the ZFC and FC curves occurs at $T_{SR} \sim 138$ K as seen in Fig. 9 a). The hysteresis loops at 100 K and the partial disappearance of the hysteresis at 50 K suggest that the Cr$^{+3}$ spins rotate from $\Gamma_4$ (G$_x$, A$_y$, F$_z$) to an incomplete $\Gamma_1$ (A$_x$, G$_y$, C$_z$) configuration since incipient hysteresis loop is observed at 50 K. It is possible that the complete $\Gamma_1$ (A$_x$, G$_y$, C$_z$) configuration could be attained just before $T^* \sim 36$ K (Fig. 9 a) as is seen, for example, in Fig. 9 b) for x=0.10 composition. There, the $T_{SR}$ increases up to $\sim 145$ K and the hysteresis loop at 100 K and the subsequent vanishing at 50 K indicate that the spin reorientation changes from $\Gamma_4$ (G$_x$, A$_y$, F$_z$) to non-magnetic $\Gamma_1$ (A$_x$, G$_y$, C$_z$) configuration at $T^* \sim 45$ K.  The result also indicates that the easy-axis of magnetization rotates beginning at $T_{SR}$ and moving continuously with decreasing temperature, to finally finish at lower temperatures (G$_x$ $\rightarrow$ G$_y$), $T \sim T^*$.

\begin{figure*}
\includegraphics[scale=0.68]{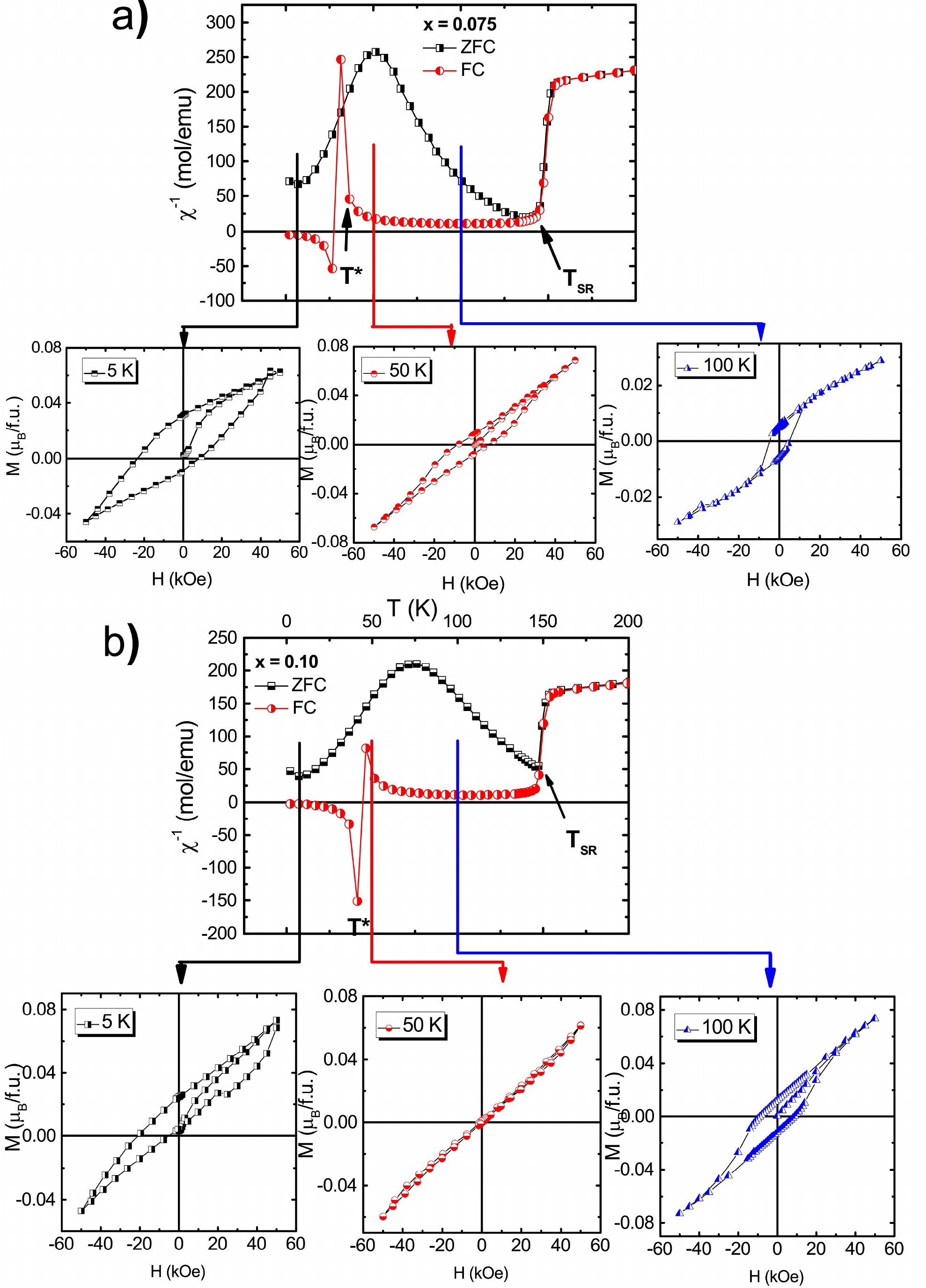}
\caption{The $\chi^{-1} - T$ data in the ZFC and FC mode along with the $M-H$ data at 5, 50 and 100 are shown for a) x= 0.075 and b) 0.10. $T_{SR}$ indicate temperature where occurs the spin reorientation and $T^*$ the compensation temperature, temperature.}
\end{figure*}

According to L. M. Levison et al.\cite{levison69}, the easy axis either rotates continuously from $T_{SR}$ down toward lower temperature ($T_2$) or jumps discontinuously with hysteresis effects. T. Yamaguchi \cite{yamaguchi74} showed that the antisymmetric (D-M) and the anisotropic-symmetric exchange interaction between M$^{+3}$ and R$^{+3}$ spins are responsible for both, the rotational SR and the abrupt SR transition, generally occurring between $T_N$ and the magnetic transition of the R$^{+3}$ sublattice. Note that the weak ferromagnetism vanishes into G$_y$-AFM (F$_z$=0) when the magnetization reversal takes place at $T^*$, contrary to what happens in \ce{NdCrO_3} where a jump in the total specific heat confirms a first order transition as a consequence of a sudden spin rotation at $T_{SR}$. The single anisotropic ion of Nd$^{+3}$ is responsible for the easy axis rotation \cite{bartolome00}. The abrupt spin rotation, in many cases, changes the reversal magnetization to positive magnetization values such as is observed in \ce{GdCrO_3} and \ce{TmCrO_3} compounds \cite{yoshii01a,yoshii12}. Two explanations of these results are plausible in \ce{Y_{1-x}Pr_xCrO_3} with $0.025 \leq x \leq  0.10$ composition: i) the exchange interaction between Pr$^{+3}$-Cr$^{+3}$ ions play a crucial role in inducing SR and ii) the gradual disappearance of the hysteresis loops when cooling from $T_{SR}$ to lower temperatures (i. e., 50 K) suggest that the spin rotates continuously from $T_{SR}$ down to $T \sim T^*$. Here, from the magnetization results, we infer that the spin reorientation begins at $T_{SR}$ with $\Gamma_4$ (G$_x$, A$_y$, F$_z$) configuration and finish at $T \sim T^*$ with $\Gamma_1$ (A$_x$, G$_y$, C$_z$) configuration for x=0.05, 0.075 and 0.10 compositions.

\begin{figure}[h]
\begin{center}
\includegraphics[scale=0.4]{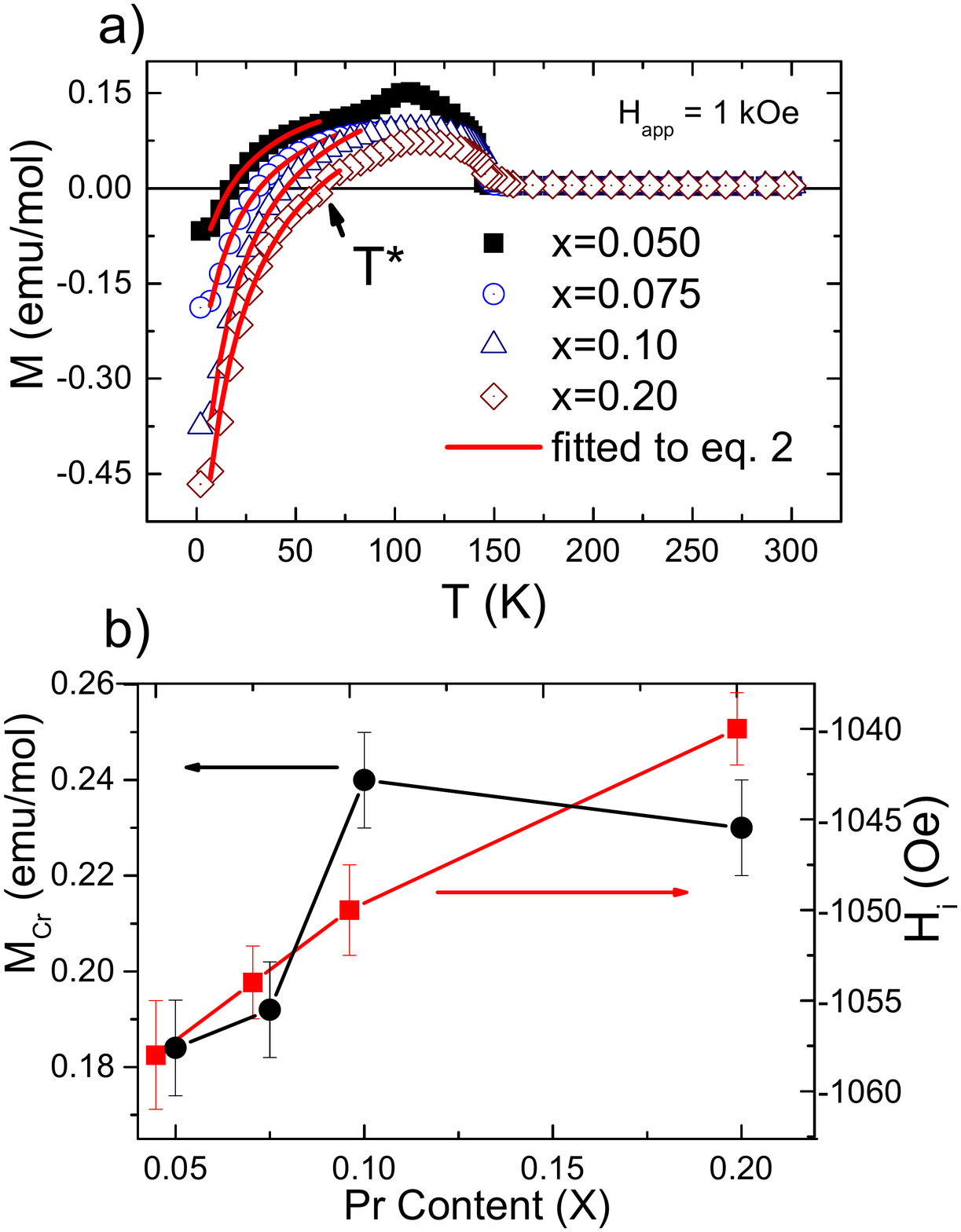}
\caption{a) Susceptibility curves in the FC mode for \ce{Y_{1-x}Pr_xCrO_3} with $0.050\leq x\leq 0.20$ content b) The Cr magnetization and the $H_i$ obtained from fitting in the same range of composition.}
\end{center}
\end{figure}

The magnetization reversal below $T^*$ is another complex magnetic behavior occurring in \ce{Y_{1-x}Pr_xCrO_3} between x=0.05 and 0.2 Pr composition. After cooling in an applied magnetic field through $T_N$ ($\sim 140$ K), the Cr$^{+3}$ sublattice imposes an internal field ($H_i$) affecting the Pr magnetic moments. Fig. 10 a) shows the FC measurements over a whole range of temperatures, under an applied field of 1 kOe for $0.05 \leq x \leq 0.20$ samples. As temperature decreases, the FC curves show a maximum and then the diamagnetic signal occurs at $T^*$. The appearance of $H_i$ can be explained according to A. H. Cooke model \cite{cooke74}. Here, the diamagnetic behavior can be fitting using the following formula:
\begin{equation}
M = M_{Cr} + \frac{C_{Pr}(H_i + H_{ap})}{(T + \theta)}
\end{equation}

The equation describes the total magnetization of the system assuming two magnetic sublattices, one formed by Cr$^{3+}$ and the other by Pr$^{3+}$ ions. There, $M_{cr}$ is the saturation magnetization of the Cr$^{3+}$ sublattice, the second term follows the Curie-Weiss law and it is associated with the paramagnetic contribution of Pr$^{3+}$ ions. The $H_i$ and $H_{app}$ are the internal field and the applied magnetic field, respectively. Furthermore, the $\theta_c$ is the Weiss temperature and the $C_{Pr} = xC_{exp}$ is attributed to the experimental Curie constant that depends on Pr$^{3+}$ composition, x. The FC curves were fitted at low temperatures and the fitting range are shown by the solid line in Fig. 10 a). The values of the $M_{Cr}$ and the $H_i$ as a function of Pr content are plotted in Fig. 10 b). We can see a slight increases in the $M_{Cr}$ taking values of 0.18 - 0.24 emu/mol between 0.075 - 0.1 compositions with a slight decrease at 0.22 emu/mol for x=0.20. The $H_i$ also slightly decreases from -1058 to -1040 Oe with increasing Pr ion composition.  The values for $\theta_c$ are between -20 and -17 K. These results contrast with those obtained in other orthocromite systems; for example, while negative magnetization in \ce{La_{1-x}Pr_xCrO_3} occurs in a wider range of Pr concentration \cite{yoshii01} ($0.2 \leq x \leq 0.8$), in \ce{Y_{1-x}Pr_xCrO_3} occurs in a narrower Pr concentration ($0.05 \leq x \leq 0.2$). Furthermore, the $M_{Cr}$ values obtained from fitting at 1 kOe are smaller than those obtained for other homologous compounds. For example, values of 40 emu/mol and $H_i \sim 8500$ Oe were reported in the \ce{La_{0.2}Pr_{0.8}CrO_3} compound \cite{yoshii01}. For \ce{La_{0.5}Gd_{0.5}CrO_3} magnetization values of 53 emu/mol and Hi ~828 Oe at 500 Oe were reported \cite{sharma10}. For \ce{NdCr_{1-x}Fe_xO_3} magnetization values from $\sim 7.5$ to $\sim 23$ emu/mol for $0.05 \leq x \leq 0.20$ and 1 kOe were reported \cite{bora15}. The low magnetization values imply that the ferromagnetic signal from the Cr sublattice is weak below $T^*$ increasing slightly with Pr$^{+3}$ content. Below $T_N$, the weak ferromagnetic moments of the Cr-sublattice impose a local internal field over the Pr$^{+3}$ moments. In this case, we assume that the Pr moments are randomly distributed at the $4c$ crystallographic site, namely the long-range ordering of the Pr-Pr sublattice is discarded. Note in Fig. 10 b) that the hysteresis curve for x=0.1 vanishes just before $T^*$ ($\sim 50$ K). At this point the $M_{Cr}$ contribution is almost zero because the spin rotates from the magnetic configuration to non-magnetic configuration,  $\Gamma_4$ (G$_x$, A$_y$, F$_z$) $\rightarrow$ $\Gamma_1$ (A$_x$, G$_y$, C$_z$). The induced local internal field ($-H_i$) at the Pr site, under further cooling, exceeds the applied field ($H_{appl}$) just below $T^*$, so that under applied field the total magnetization becomes negative since the Pr moments locally increase with decreasing temperature. The result also implies that below $T^*$, a new AFM spin configuration with a low magnetization ($M_{Cr} \sim 0.20$ emu/mol) takes place due to stronger interaction between magnetic moments of Pr with Cr ions inducing a new magnetic, $\Gamma_2$ (F$_x$, C$_y$, G$_z$) configuration. This fact could explain the lower $M_{Cr}$ values obtained from the fitting below of $T^*$ with respect to that reported in other compounds. The ferromagnetic component in $\Gamma_2$ (F$_x$) configuration below of $T^*$ in the \ce{Y_{1-x}Pr_xCrO_3} compound is weaker than that observed in, for example, \ce{GdCrO_3} where magnetization values, $M_{Cr} = 100-400$ emu/mol in  $\Gamma_4$ (F$_z$) configuration were obtained \cite{yoshii01a,cooke74}.

\begin{figure}[t]
\begin{center}
\includegraphics[scale=0.4]{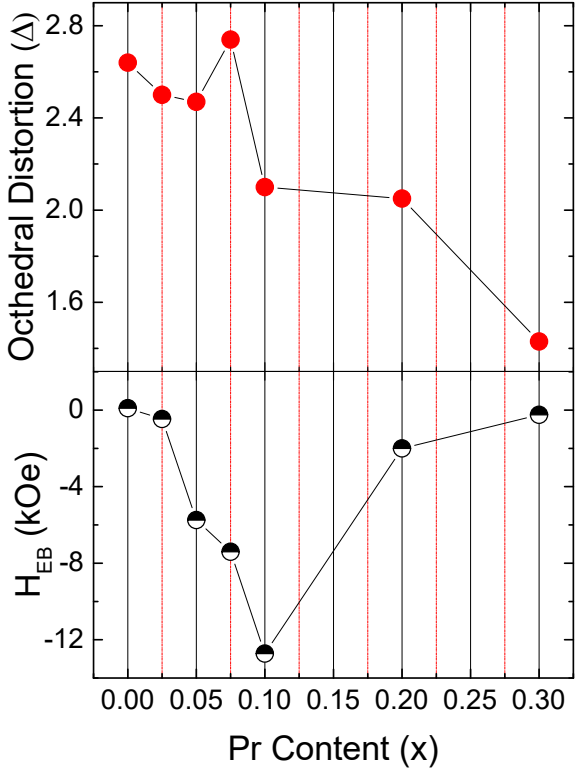}
\caption{The octahedral distortion ($\Delta$) behavior at RT is compared with the effective moments in the paramagnetic state ($\mu_{eff}$) and the exchange-bias field ($H_{EB}$) at 5 K vs Pr content. }
\end{center}
\end{figure}

A different situation occurs at lower temperature ($< T^*$). The hysteresis loops show that the non-magnetic $\Gamma_1$ (A$_x$, G$_y$, C$_z$) configuration is not maintained at 5 K.  We observe three important characteristics in the hysteresis curves at 5 K for \ce{Y_{1-x}Pr_xCrO_3} with $0.025  \leq x \leq 0.3$: i) the magnetic moment of the Cr$^{+3}$ ions continues to rotate below $T^*$. ii) the $H_c$ and $M_r$ show an anomalous behavior for compositions between 0.075-0.1 compositions. iii) negative exchange bias in the whole range of Pr$^{+3}$ compositions. iv) the magnetization and specific heat curves show an emergence of a peak at 10 K. These results seem to indicate that a new magnetic structure takes place below $T^*$ and the magnetization results are consistent for the coexistence of FM and AFM domains, which explains the exchange bias effect for \ce{Y_{1-x}Pr_xCrO_3} for x=0.025-0.30 composition. Recently, D. Deng et al. \cite{deng15} performed neutron diffraction studies on \ce{Y_{0.9}Pr_{0.1}CrO_3} finding that the $\Gamma_2$ (F$_x$, C$_y$, G$_z$) configuration is dominant and it is associated with the Cr-Cr exchange interaction at 3 K. Here, the hysteresis curves of Fig. 5 d) confirm the development of the $\Gamma_2$ (F$_x$, C$_y$, G$_z$) spin configuration at 5 K in agreement with the results of D. Deng et al \cite{deng15}. The result implies that the $\Gamma_4$ (G$_x$, A$_y$, F$_z$) $\rightarrow$ $\Gamma_1$ (A$_x$, G$_y$, C$_z$) $\rightarrow$ $\Gamma_2$ (F$_x$, C$_y$, G$_z$) spins configuration (Cr$^{+3}$-spin) take place from $T_N$ to 5 K for x = 0.075 - 0.1 of Pr$^{+3}$ content. We have to mention above that the non-magnetic $\Gamma_1$ (A$_x$, G$_y$, C$_z$) spin configuration takes place at $T^*$. With further decreasing temperature, the anisotropy of Pr$^{+3}$ ions overcome those of the Cr$^{+3}$ ions and cause the Cr$^{+3}$ moments to rotate continuously toward the $\Gamma_2$ (F$_x$, C$_y$, G$_z$) spin configuration. It is worthwhile noting two composition regions where the magnetization is controlled by the crystal anisotropy (tilting and octahedral distortion) and the other by a possible short order Pr-Pr exchange interaction increasing the total magnetization with increasing Pr ion content.  To justify these scenarios, we investigate the octahedral distortion effect on the magnetization taking into consideration $H_{EB}$ values at 5 K as a function of Pr content as is displayed in Fig. 11. A clear tuning of the octahedral distortion and the $H_{EB}$ values from 0.05 to 0.1 of Pr content is observed. There, a sudden decrease of the octahedral distortion occurring at 0.1 of Pr content accompanied with a maximum in the negative $H_{EB}$ values is observed. Zhou et al. \cite{zhou08} showed that the orthorhombic ($Pbnm$ S. G.) crystalline structure is not rigid and the R substitution causes the cooperative-site rotation inducing an intrinsic octahedral distortion; accordingly, the $t-e$ hybridization should be inevitable. It has been observed that the cooperative-site rotation shifts $T_N$ and the hybridization are the main responsible affecting the Cr-O-Cr exchange interaction \cite{zhou10}. Thus, we infer that the $t-e$ hybridization of the Cr-O-Cr is modified by the magnetic moment of the Pr ion inducing the anomalous behavior in the octahedral distortion, and thus influencing directly the spins configuration, the magnetization and the negative exchange bias effect below $T_N$ as is clearly seen in Fig. 11 a, b). This is another important factor that we must not ignore, which also plays an important role in the development of these complex dynamics magnetic phenomena since the octahedral distortion is governed by the octahedral tilting along the [110] and [001] directions and this, in turn, tune the magnetic properties. The experimental evidence that supports this fact come from the magnetic exchange bias field since these phenomena imply the presence of ferromagnetic domains coexisting with AFM domains, both coming from an independent nature. Note that the strong exchange anisotropy is more visible when the octahedral distortion attains a maximum at 0.1 Pr composition. In other words, the exchange interaction between FM and AFM domains is tuned by octahedral distortion for $0.025 \leq x \leq 0.1$ as seen in Fig. 11. On the other hand, the increases of the $H_c$ and $M_r$ for higher Pr concentration (x=0.2 and 0.3) should come from another source.

Finally, the development of a peak at 10 K strongly suggest that is related to the Pr - Pr exchange interaction, which becomes visible at low temperature. For \ce{RCrO_3}, a second magnetic transition at lower temperatures ($T < T_N$) has been associated with the magnetic R$^{+3}$ ions. The R$^{+3}$- R$^{+3}$ exchange interaction is the weakest interaction and occurs at $T \leq 10$ K \cite{yamaguchi74}. Several experimental pieces of evidence in the present study predict a new magnetic configuration presumably due to Pr$^{+3}$ (S=1) spins, which could be ferromagnetic. For example, the magnetic contribution to the specific heat showed a peak at about 10 K, increasing continuously as the Pr$^{+3}$ content increased (see inset in Fig. 8b). Other evidence arises from ZFC magnetization curves (arrows in Fig. 4 c-f) displaying an incipient peak at about 10 K, which increases in magnitude with increasing Pr content. Thus, for example, it is known that for the \ce{PrCrO_3} compound, $T_N$ occurs at $\sim 238$ K. The spontaneous magnetization of the Cr$^{+3}$ ions is along the [100] direction, taking the $\Gamma_2$ (F$_x$, C$_y$, G$_z$) spin configuration. However, the magnetization data do not show any feature related with to the second magnetic transition below of 20 K in \ce{PrCrO_3} \cite{sardar11,gordon76}.  On the other hand, neutron diffraction studies have failed to resolve the second magnetic transition as well as the spin configuration of the Pr-Pr exchange interactions at lower temperatures. E. F. Bertaut et al. \cite{bertaut66} observed spin ordering with a weak ferroelectric component, F$_x$, above 4.2 K without referring to a Pr magnetic ordering. Afterwards, N. Shamir et al. \cite{shamir81} identified a weak magnetic structure (C$_y$F$_x$ magnetic structure) by neutron scattering studies at 9 K for the \ce{PrCrO_3} compound. Recently, D. Deng et al. \cite{deng15} performed neutron diffraction studies in \ce{Y_{0.9}Pr_{0.1}CrO_3} and they did not find signs of magnetic peaks related to the Pr-Pr exchange interaction at 3 K, presumably due to insufficient resolution. However, we discard the assumption of long-range Pr-Pr exchange interaction in \ce{Y_{1-x}Pr_xCrO_3} compound since the results in the paramagnetic regime (not shown) did not show an anomaly at about 238 K that indicate a chemical phase separation of the  \ce{PrCrO_3} ($T_N \sim 238$ K) into the \ce{YCrO_3} ($T_N \sim 142$ K) matrix. This fact also suggests that the Pr-ions are chemically disordered in the 4c site. Three plausible answers could explain the magnetic behavior at low temperature: (i) the magnetic moments of the Pr ions ($f^2$, S=1) is frozen, at the crystallographic $4c$ site, inducing spin-glass clusters at about 10 K (ii) the octahedral distortion promotes the formation of ferromagnetic domains around Pr sites.  (iii) exchange interaction between DM and the single ion magnetic anisotropy of the Pr ion \cite{bora15,dong09}. The DM interaction is responsible for WFM domains and the single ion magnetic anisotropy arising from the random magnetic moment of the Pr ion ($4f^2$ - S=1) could be responsible of the ferromagnetic domains. The net result is an increase of the ferromagnetic domains as Pr increased in the \ce{YCrO_3} compound.

\begin{figure}
\begin{center}
\includegraphics[scale=0.37]{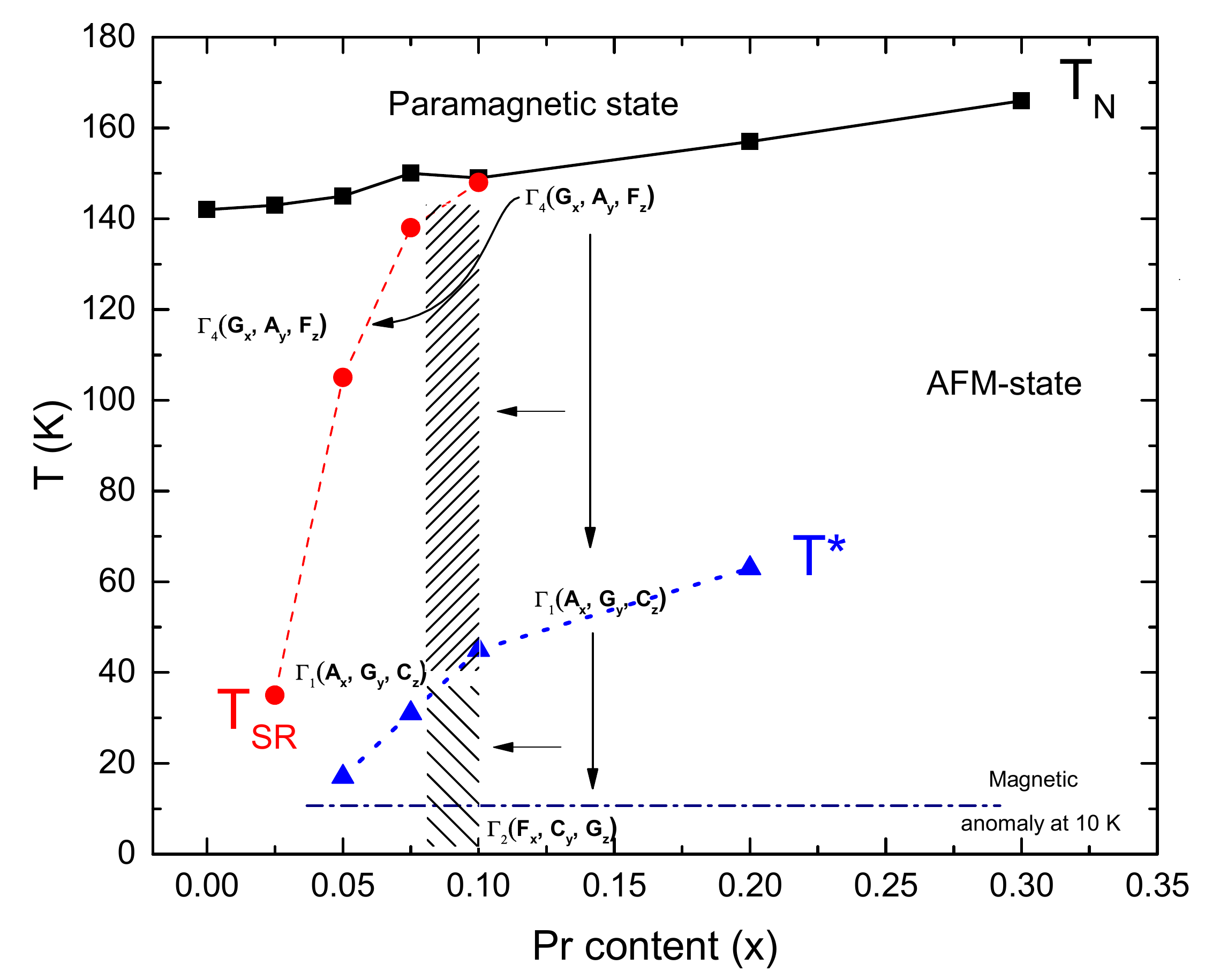}
\caption{Phase diagram of \ce{Y_{1-x}Pr_xCrO_3} ($0 \leq x \leq 0.3$). The solid, and dash line for $T_N$, $T_{SR}$ and $T^*$ are guides to the eye. There, $T_N$, $T_{SR}$ and $T^*$ denote the N\'eel, spin reorientation and compensation temperatures, respectively. The hatched zone denotes the spin dynamic configuration from $\Gamma_4$ (G$_x$, A$_y$, F$_z$) $\rightarrow$ $\Gamma_1$ (A$_x$, G$_y$, C$_z$) $\rightarrow$ $\Gamma_2$ (F$_x$, C$_y$, G$_z$) as a function of temperature.}
\end{center}
\end{figure}

Based on magnetization and specific heat results; it was possible to propose a phase diagram that summarizes the main result obtained in \ce{Y_{1-x}Pr_xCrO_3} ($0  \leq x \leq 0.3$) as seen in Fig. 12. The solid squares represent the N\'eel temperature and the boundary between the paramagnetic and the AFM state. The solid red circles show the spin reorientation transition which is increased from 30 K for x=0.025 to $\sim T_N$ for x=0.01 composition. At lower temperatures, the compensation temperature, $T^*$ appears at about 35 K for x=0.05 and increases up to 63 K for x=0.20 (solid blue triangle). At 10 K (dash-dot line), the magnetic and specific heat studies show an anomaly that does not change in temperature with increasing Pr$^{+3}$ content and that is presumably due to the spin-freeze of the Pr$^{+3}$ ion. A possible short-range ordering of praseodymium is discarded since there is no evidence of phase separation of \ce{YCrO_3} and \ce{PrCrO_3} compound. In addition, the experimental evidence shows a complex spin dynamics ground state configuration below $T_N$. We found below $T_{SR}$ that the easy magnetization axis rotates continuously from one spin configuration to another at lower temperatures. For example, for x=0.10, the spin reorientation occurs at $T_{SR} \sim 148$ K (hatched zone). There, the $\Gamma_4$ (G$_x$, A$_y$, F$_z$) AFM structure rotates continuously ending at $\Gamma_1$ (A$_x$, G$_y$, C$_z$) configuration close to $T^*$. Afterward, $\Gamma_1$ (A$_x$, G$_y$, C$_z$) $\rightarrow$ $\Gamma_2$ (F$_x$, C$_y$, G$_z$) spin configuration takes place as is seen by the magnetic hysteresis at 5 K (see Fig. 5 c). Recently,  L. Bellaiche et al \cite{bellaiche12}. obtained a simple law that governs the magnetic coupling and takes into account the tilting of the oxygen octahedra  (antiferrodistortive quantity, $w_i$) and Dzyaloshinsky-Moriya interaction. The law predicts that the inclination of the octahedrons relaxes the magnetic structure, and can thus adopt secondary magnetic orders (as $\Gamma_1$ and $\Gamma_2$ spin configuration in this study) in order to minimize the total energy.

Though these last assertions are not conclusive, it is worthwhile mentioning that the experimental magnetic behavior as well as the specific heat studies are a good approach to investigate the spin dynamic induced by temperature; however, a deeper look by neutron diffraction studies is recommended to confirm the spin ground state configuration in \ce{Y_{1-x}Pr_xCrO_3}.

\section{Conclusion}
In summary, we present a systematic study of the structure and electronic properties of the YCrO3 doped with Pr. We begin from a detailed crystalline characterization followed by the studies of the magnetic and thermal properties of \ce{Y_{1-x}Pr_xCrO_3} with $0 \leq x \leq 0.3$. We found that the magnetic properties are strongly coupled to the structural parameters. Spin reorientation, magnetization reversal and exchange bias effects induced by temperature in a short range of Pr composition is a consequence of the competition of antisymmetric (D-M) and anisotropic-symmetric exchange interaction between Pr$^{+3}$-Cr$^{+3}$ ions. It is found that not only the D-M and the single anisotropic ion of Pr magnetic moments are responsible for the complex magnetic behavior below $T_N$ but also the octahedral distortion. The second order transition of the spin reorientation is dynamic because the easy-axis of magnetization rotates beginning at $T_{SR}$ and moving continuously with decreasing temperature, following the $\Gamma_4$ (G$_x$, A$_y$, F$_z$) $\rightarrow$ $\Gamma_1$ (A$_x$, G$_y$, C$_z$) $\rightarrow$ $\Gamma_2$ (F$_x$, C$_y$, G$_z$) spin configuration (Cr-spin) from $T_N \rightarrow T^* \rightarrow 5$ K. Furthermore, this fact not only suggests the energetic condition for the presence of both phenomena but also the development of ferromagnetic domains coexisting with antiferromagnetic domains, both coming from independent sources. The close coexistence of both domains induces the development of the exchange magnetic anisotropy at lower temperatures. There is not a signal of long-range Pr-Pr interaction at low temperature since the Pr-ion is randomly distributed in the crystal lattice. However, the magnetization and specific heat measurements reveal a hump around 10 K, which infer spin-glass of the Pr moments.


%



\begin{acknowledgments}
A. Dur\'an. thanks to project No. IN103016 (PAPITT-UNAM), R. Escamilla to projyect No. IN106116/28 (PAPITT-UNAM) and R. Escudero to project No. IT100217 (PAPITT-UNAM). E. Verd\'in thanks to the intercambio acad\'emico UNISON-UNAM (CNYN). Specials thanks to Dr. J. Siqueiros by the fruitful comments and the proofreading our manuscript. The technical assistance of P. Casillas, I. Gradilla, E. Aparicio, A. Pompa (IIM), A. L\'opez (IIM) and A. Tejeda-Cruz (IIM) is acknowledged.
\end{acknowledgments}

\bibliography{Duran}

\begin{thebibliography}{59}%
\makeatletter
\providecommand \@ifxundefined [1]{%
 \@ifx{#1\undefined}
}%
\providecommand \@ifnum [1]{%
 \ifnum #1\expandafter \@firstoftwo
 \else \expandafter \@secondoftwo
 \fi
}%
\providecommand \@ifx [1]{%
 \ifx #1\expandafter \@firstoftwo
 \else \expandafter \@secondoftwo
 \fi
}%
\providecommand \natexlab [1]{#1}%
\providecommand \enquote  [1]{``#1''}%
\providecommand \bibnamefont  [1]{#1}%
\providecommand \bibfnamefont [1]{#1}%
\providecommand \citenamefont [1]{#1}%
\providecommand \href@noop [0]{\@secondoftwo}%
\providecommand \href [0]{\begingroup \@sanitize@url \@href}%
\providecommand \@href[1]{\@@startlink{#1}\@@href}%
\providecommand \@@href[1]{\endgroup#1\@@endlink}%
\providecommand \@sanitize@url [0]{\catcode `\\12\catcode `\$12\catcode
  `\&12\catcode `\#12\catcode `\^12\catcode `\_12\catcode `\%12\relax}%
\providecommand \@@startlink[1]{}%
\providecommand \@@endlink[0]{}%
\providecommand \url  [0]{\begingroup\@sanitize@url \@url }%
\providecommand \@url [1]{\endgroup\@href {#1}{\urlprefix }}%
\providecommand \urlprefix  [0]{URL }%
\providecommand \Eprint [0]{\href }%
\providecommand \doibase [0]{http://dx.doi.org/}%
\providecommand \selectlanguage [0]{\@gobble}%
\providecommand \bibinfo  [0]{\@secondoftwo}%
\providecommand \bibfield  [0]{\@secondoftwo}%
\providecommand \translation [1]{[#1]}%
\providecommand \BibitemOpen [0]{}%
\providecommand \bibitemStop [0]{}%
\providecommand \bibitemNoStop [0]{.\EOS\space}%
\providecommand \EOS [0]{\spacefactor3000\relax}%
\providecommand \BibitemShut  [1]{\csname bibitem#1\endcsname}%
\let\auto@bib@innerbib\@empty
\bibitem [{\citenamefont {Salamon}\ and\ \citenamefont
  {Jaime}(2001)}]{salamon01}%
  \BibitemOpen
  \bibfield  {author} {\bibinfo {author} {\bibfnamefont {M.}~\bibnamefont
  {Salamon}}\ and\ \bibinfo {author} {\bibfnamefont {M.}~\bibnamefont
  {Jaime}},\ }\href@noop {} {\bibfield  {journal} {\bibinfo  {journal} {Reviews
  of Modern Physics}\ }\textbf {\bibinfo {volume} {73}},\ \bibinfo {pages}
  {583} (\bibinfo {year} {2001})}\BibitemShut {NoStop}%
\bibitem [{\citenamefont {Kimura}\ \emph {et~al.}(2003)\citenamefont {Kimura},
  \citenamefont {Goto}, \citenamefont {Shintani}, \citenamefont {Ishizaka},
  \citenamefont {Arima},\ and\ \citenamefont {Tokura}}]{kimura03}%
  \BibitemOpen
  \bibfield  {author} {\bibinfo {author} {\bibfnamefont {T.}~\bibnamefont
  {Kimura}}, \bibinfo {author} {\bibfnamefont {T.}~\bibnamefont {Goto}},
  \bibinfo {author} {\bibfnamefont {H.}~\bibnamefont {Shintani}}, \bibinfo
  {author} {\bibfnamefont {K.}~\bibnamefont {Ishizaka}}, \bibinfo {author}
  {\bibfnamefont {T.}~\bibnamefont {Arima}}, \ and\ \bibinfo {author}
  {\bibfnamefont {Y.}~\bibnamefont {Tokura}},\ }\href@noop {} {\bibfield
  {journal} {\bibinfo  {journal} {Nature}\ }\textbf {\bibinfo {volume} {426}},\
  \bibinfo {pages} {55} (\bibinfo {year} {2003})}\BibitemShut {NoStop}%
\bibitem [{\citenamefont {Cheong}\ and\ \citenamefont
  {Mostovoy}(2007)}]{cheong07}%
  \BibitemOpen
  \bibfield  {author} {\bibinfo {author} {\bibfnamefont {S.-W.}\ \bibnamefont
  {Cheong}}\ and\ \bibinfo {author} {\bibfnamefont {M.}~\bibnamefont
  {Mostovoy}},\ }\href@noop {} {\bibfield  {journal} {\bibinfo  {journal}
  {Nature Materials}\ }\textbf {\bibinfo {volume} {6}},\ \bibinfo {pages} {13}
  (\bibinfo {year} {2007})}\BibitemShut {NoStop}%
\bibitem [{\citenamefont {Eerenstein}\ \emph {et~al.}(2006)\citenamefont
  {Eerenstein}, \citenamefont {Mathur},\ and\ \citenamefont
  {Scott}}]{eerenstein06}%
  \BibitemOpen
  \bibfield  {author} {\bibinfo {author} {\bibfnamefont {W.}~\bibnamefont
  {Eerenstein}}, \bibinfo {author} {\bibfnamefont {N.}~\bibnamefont {Mathur}},
  \ and\ \bibinfo {author} {\bibfnamefont {J.}~\bibnamefont {Scott}},\
  }\href@noop {} {\bibfield  {journal} {\bibinfo  {journal} {Nature}\ }\textbf
  {\bibinfo {volume} {422}},\ \bibinfo {pages} {759} (\bibinfo {year}
  {2006})}\BibitemShut {NoStop}%
\bibitem [{\citenamefont {Rajeswaran}\ \emph {et~al.}(2012)\citenamefont
  {Rajeswaran}, \citenamefont {Khomskii}, \citenamefont {Zvezdin},
  \citenamefont {Rao},\ and\ \citenamefont {Sundaresan}}]{rajeswaran12}%
  \BibitemOpen
  \bibfield  {author} {\bibinfo {author} {\bibfnamefont {B.}~\bibnamefont
  {Rajeswaran}}, \bibinfo {author} {\bibfnamefont {D.}~\bibnamefont
  {Khomskii}}, \bibinfo {author} {\bibfnamefont {A.}~\bibnamefont {Zvezdin}},
  \bibinfo {author} {\bibfnamefont {C.}~\bibnamefont {Rao}}, \ and\ \bibinfo
  {author} {\bibfnamefont {A.}~\bibnamefont {Sundaresan}},\ }\href@noop {}
  {\bibfield  {journal} {\bibinfo  {journal} {Physical Review B}\ }\textbf
  {\bibinfo {volume} {86}},\ \bibinfo {pages} {214409} (\bibinfo {year}
  {2012})}\BibitemShut {NoStop}%
\bibitem [{\citenamefont {Kumar}\ and\ \citenamefont {Yusuf}(2015)}]{kumar15}%
  \BibitemOpen
  \bibfield  {author} {\bibinfo {author} {\bibfnamefont {A.}~\bibnamefont
  {Kumar}}\ and\ \bibinfo {author} {\bibfnamefont {S.}~\bibnamefont {Yusuf}},\
  }\href@noop {} {\bibfield  {journal} {\bibinfo  {journal} {Physics Reports}\
  }\textbf {\bibinfo {volume} {556}},\ \bibinfo {pages} {1} (\bibinfo {year}
  {2015})}\BibitemShut {NoStop}%
\bibitem [{\citenamefont {Bellaiche}\ \emph {et~al.}(2012)\citenamefont
  {Bellaiche}, \citenamefont {Gui},\ and\ \citenamefont
  {Kornev}}]{bellaiche12}%
  \BibitemOpen
  \bibfield  {author} {\bibinfo {author} {\bibfnamefont {L.}~\bibnamefont
  {Bellaiche}}, \bibinfo {author} {\bibfnamefont {Z.}~\bibnamefont {Gui}}, \
  and\ \bibinfo {author} {\bibfnamefont {I.}~\bibnamefont {Kornev}},\
  }\href@noop {} {\bibfield  {journal} {\bibinfo  {journal} {Journal of
  Physics: Condensed Matter}\ }\textbf {\bibinfo {volume} {24}},\ \bibinfo
  {pages} {312201} (\bibinfo {year} {2012})}\BibitemShut {NoStop}%
\bibitem [{\citenamefont {Ramesh}\ and\ \citenamefont
  {Spaldin}(2007)}]{ramesh07}%
  \BibitemOpen
  \bibfield  {author} {\bibinfo {author} {\bibfnamefont {R.}~\bibnamefont
  {Ramesh}}\ and\ \bibinfo {author} {\bibfnamefont {N.}~\bibnamefont
  {Spaldin}},\ }\href@noop {} {\bibfield  {journal} {\bibinfo  {journal}
  {Nature Materials}\ }\textbf {\bibinfo {volume} {6}},\ \bibinfo {pages} {21}
  (\bibinfo {year} {2007})}\BibitemShut {NoStop}%
\bibitem [{\citenamefont {Bibes}(2008)}]{bibes08}%
  \BibitemOpen
  \bibfield  {author} {\bibinfo {author} {\bibfnamefont {A.}~\bibnamefont
  {Bibes}},\ }\href@noop {} {\bibfield  {journal} {\bibinfo  {journal} {Nature
  Materials}\ }\textbf {\bibinfo {volume} {7}},\ \bibinfo {pages} {425}
  (\bibinfo {year} {2008})}\BibitemShut {NoStop}%
\bibitem [{\citenamefont {Ramesh}(2010)}]{ramesh10}%
  \BibitemOpen
  \bibfield  {author} {\bibinfo {author} {\bibfnamefont {R.}~\bibnamefont
  {Ramesh}},\ }\href@noop {} {\bibfield  {journal} {\bibinfo  {journal} {Nature
  Materials}\ }\textbf {\bibinfo {volume} {9}},\ \bibinfo {pages} {380}
  (\bibinfo {year} {2010})}\BibitemShut {NoStop}%
\bibitem [{\citenamefont {Prado-Gonjal}\ \emph {et~al.}(2013)\citenamefont
  {Prado-Gonjal}, \citenamefont {Schmidt}, \citenamefont {Romero},
  \citenamefont {\'Avila}, \citenamefont {Amador},\ and\ \citenamefont
  {Mor\'an}}]{prado13}%
  \BibitemOpen
  \bibfield  {author} {\bibinfo {author} {\bibfnamefont {J.}~\bibnamefont
  {Prado-Gonjal}}, \bibinfo {author} {\bibfnamefont {R.}~\bibnamefont
  {Schmidt}}, \bibinfo {author} {\bibfnamefont {J.}~\bibnamefont {Romero}},
  \bibinfo {author} {\bibfnamefont {D.}~\bibnamefont {\'Avila}}, \bibinfo
  {author} {\bibfnamefont {U.}~\bibnamefont {Amador}}, \ and\ \bibinfo {author}
  {\bibfnamefont {E.}~\bibnamefont {Mor\'an}},\ }\href@noop {} {\bibfield
  {journal} {\bibinfo  {journal} {Inorganic Chemistry}\ }\textbf {\bibinfo
  {volume} {52}},\ \bibinfo {pages} {313} (\bibinfo {year} {2013})}\BibitemShut
  {NoStop}%
\bibitem [{\citenamefont {Yamaguchi}\ and\ \citenamefont
  {Tsushima}(1973)}]{yamaguchi73}%
  \BibitemOpen
  \bibfield  {author} {\bibinfo {author} {\bibfnamefont {T.}~\bibnamefont
  {Yamaguchi}}\ and\ \bibinfo {author} {\bibfnamefont {K.}~\bibnamefont
  {Tsushima}},\ }\href@noop {} {\bibfield  {journal} {\bibinfo  {journal}
  {Physical Review B}\ }\textbf {\bibinfo {volume} {8}},\ \bibinfo {pages}
  {5187} (\bibinfo {year} {1973})}\BibitemShut {NoStop}%
\bibitem [{\citenamefont {Moriya}(1960)}]{moriya}%
  \BibitemOpen
  \bibfield  {author} {\bibinfo {author} {\bibfnamefont {T.}~\bibnamefont
  {Moriya}},\ }\href@noop {} {\bibfield  {journal} {\bibinfo  {journal}
  {Physical Review}\ }\textbf {\bibinfo {volume} {120}},\ \bibinfo {pages} {91}
  (\bibinfo {year} {1960})}\BibitemShut {NoStop}%
\bibitem [{\citenamefont {Treves}(1962)}]{treves}%
  \BibitemOpen
  \bibfield  {author} {\bibinfo {author} {\bibfnamefont {D.}~\bibnamefont
  {Treves}},\ }\href@noop {} {\bibfield  {journal} {\bibinfo  {journal}
  {Physical Review}\ }\textbf {\bibinfo {volume} {125}},\ \bibinfo {pages}
  {1843} (\bibinfo {year} {1962})}\BibitemShut {NoStop}%
\bibitem [{\citenamefont {Sergienko}\ and\ \citenamefont
  {Dagotto}(2006)}]{sergienko06}%
  \BibitemOpen
  \bibfield  {author} {\bibinfo {author} {\bibfnamefont {I.}~\bibnamefont
  {Sergienko}}\ and\ \bibinfo {author} {\bibfnamefont {E.}~\bibnamefont
  {Dagotto}},\ }\href@noop {} {\bibfield  {journal} {\bibinfo  {journal}
  {Physical Review B}\ }\textbf {\bibinfo {volume} {73}},\ \bibinfo {pages}
  {094434} (\bibinfo {year} {2006})}\BibitemShut {NoStop}%
\bibitem [{\citenamefont {Yoshii}\ and\ \citenamefont
  {Nakamura}(2000)}]{yoshii00}%
  \BibitemOpen
  \bibfield  {author} {\bibinfo {author} {\bibfnamefont {K.}~\bibnamefont
  {Yoshii}}\ and\ \bibinfo {author} {\bibfnamefont {A.}~\bibnamefont
  {Nakamura}},\ }\href@noop {} {\bibfield  {journal} {\bibinfo  {journal}
  {Journal of Solid State Chemistry}\ }\textbf {\bibinfo {volume} {155}},\
  \bibinfo {pages} {447} (\bibinfo {year} {2000})}\BibitemShut {NoStop}%
\bibitem [{\citenamefont {Khomchenko}\ \emph {et~al.}(2008)\citenamefont
  {Khomchenko}, \citenamefont {Troyanchuk}, \citenamefont {Szymczak},\ and\
  \citenamefont {Szymczak}}]{khomchenko08}%
  \BibitemOpen
  \bibfield  {author} {\bibinfo {author} {\bibfnamefont {V.}~\bibnamefont
  {Khomchenko}}, \bibinfo {author} {\bibfnamefont {I.}~\bibnamefont
  {Troyanchuk}}, \bibinfo {author} {\bibfnamefont {R.}~\bibnamefont
  {Szymczak}}, \ and\ \bibinfo {author} {\bibfnamefont {H.}~\bibnamefont
  {Szymczak}},\ }\href@noop {} {\bibfield  {journal} {\bibinfo  {journal}
  {Journal of Materials Science}\ }\textbf {\bibinfo {volume} {43}},\ \bibinfo
  {pages} {5662} (\bibinfo {year} {2008})}\BibitemShut {NoStop}%
\bibitem [{\citenamefont {Mao}\ \emph {et~al.}(2011)\citenamefont {Mao},
  \citenamefont {Sui}, \citenamefont {Zhang}, \citenamefont {Su}, \citenamefont
  {Wang}, \citenamefont {Liu}, \citenamefont {Wang}, \citenamefont {Zhu},
  \citenamefont {Wang}, \citenamefont {Liu},\ and\ \citenamefont
  {Tang}}]{mao11}%
  \BibitemOpen
  \bibfield  {author} {\bibinfo {author} {\bibfnamefont {J.}~\bibnamefont
  {Mao}}, \bibinfo {author} {\bibfnamefont {Y.}~\bibnamefont {Sui}}, \bibinfo
  {author} {\bibfnamefont {X.}~\bibnamefont {Zhang}}, \bibinfo {author}
  {\bibfnamefont {Y.}~\bibnamefont {Su}}, \bibinfo {author} {\bibfnamefont
  {X.}~\bibnamefont {Wang}}, \bibinfo {author} {\bibfnamefont {Z.}~\bibnamefont
  {Liu}}, \bibinfo {author} {\bibfnamefont {Y.}~\bibnamefont {Wang}}, \bibinfo
  {author} {\bibfnamefont {R.}~\bibnamefont {Zhu}}, \bibinfo {author}
  {\bibfnamefont {Y.}~\bibnamefont {Wang}}, \bibinfo {author} {\bibfnamefont
  {W.}~\bibnamefont {Liu}}, \ and\ \bibinfo {author} {\bibfnamefont
  {J.}~\bibnamefont {Tang}},\ }\href@noop {} {\bibfield  {journal} {\bibinfo
  {journal} {Applied Physics Letters}\ }\textbf {\bibinfo {volume} {98}},\
  \bibinfo {pages} {192510} (\bibinfo {year} {2011})}\BibitemShut {NoStop}%
\bibitem [{\citenamefont {Gorter}\ and\ \citenamefont
  {Schulkes}(1953)}]{gorter53}%
  \BibitemOpen
  \bibfield  {author} {\bibinfo {author} {\bibfnamefont {E.}~\bibnamefont
  {Gorter}}\ and\ \bibinfo {author} {\bibfnamefont {J.}~\bibnamefont
  {Schulkes}},\ }\href@noop {} {\bibfield  {journal} {\bibinfo  {journal}
  {Physical Review}\ }\textbf {\bibinfo {volume} {90}},\ \bibinfo {pages} {487}
  (\bibinfo {year} {1953})}\BibitemShut {NoStop}%
\bibitem [{\citenamefont {Pauthenet}(1958)}]{pauthenet58}%
  \BibitemOpen
  \bibfield  {author} {\bibinfo {author} {\bibfnamefont {R.}~\bibnamefont
  {Pauthenet}},\ }\href@noop {} {\bibfield  {journal} {\bibinfo  {journal}
  {Journal of Applied Physics}\ }\textbf {\bibinfo {volume} {29}},\ \bibinfo
  {pages} {253} (\bibinfo {year} {1958})}\BibitemShut {NoStop}%
\bibitem [{\citenamefont {Ren}\ \emph {et~al.}(1998)\citenamefont {Ren},
  \citenamefont {Palstra}, \citenamefont {Khomskii}, \citenamefont {Pellegrin},
  \citenamefont {Nugroho}, \citenamefont {Menovsky},\ and\ \citenamefont
  {Sawatzky}}]{ren98}%
  \BibitemOpen
  \bibfield  {author} {\bibinfo {author} {\bibfnamefont {Y.}~\bibnamefont
  {Ren}}, \bibinfo {author} {\bibfnamefont {T.}~\bibnamefont {Palstra}},
  \bibinfo {author} {\bibfnamefont {D.}~\bibnamefont {Khomskii}}, \bibinfo
  {author} {\bibfnamefont {E.}~\bibnamefont {Pellegrin}}, \bibinfo {author}
  {\bibfnamefont {A.}~\bibnamefont {Nugroho}}, \bibinfo {author} {\bibfnamefont
  {A.}~\bibnamefont {Menovsky}}, \ and\ \bibinfo {author} {\bibfnamefont
  {G.}~\bibnamefont {Sawatzky}},\ }\href@noop {} {\bibfield  {journal}
  {\bibinfo  {journal} {Nature}\ }\textbf {\bibinfo {volume} {396}},\ \bibinfo
  {pages} {441} (\bibinfo {year} {1998})}\BibitemShut {NoStop}%
\bibitem [{\citenamefont {Ohkoshi}\ \emph {et~al.}(1997)\citenamefont
  {Ohkoshi}, \citenamefont {Iyoda}, \citenamefont {Fujishima},\ and\
  \citenamefont {Hashimoto}}]{ohkoshi97}%
  \BibitemOpen
  \bibfield  {author} {\bibinfo {author} {\bibfnamefont {S.}~\bibnamefont
  {Ohkoshi}}, \bibinfo {author} {\bibfnamefont {T.}~\bibnamefont {Iyoda}},
  \bibinfo {author} {\bibfnamefont {A.}~\bibnamefont {Fujishima}}, \ and\
  \bibinfo {author} {\bibfnamefont {K.}~\bibnamefont {Hashimoto}},\ }\href@noop
  {} {\bibfield  {journal} {\bibinfo  {journal} {Physical Review B}\ }\textbf
  {\bibinfo {volume} {56}},\ \bibinfo {pages} {11642} (\bibinfo {year}
  {1997})}\BibitemShut {NoStop}%
\bibitem [{\citenamefont {Ren}\ \emph {et~al.}(2000)\citenamefont {Ren},
  \citenamefont {Palstra}, \citenamefont {Khomskii}, \citenamefont {Nugroho},
  \citenamefont {Menovsky},\ and\ \citenamefont {Sawatzky}}]{ren00}%
  \BibitemOpen
  \bibfield  {author} {\bibinfo {author} {\bibfnamefont {Y.}~\bibnamefont
  {Ren}}, \bibinfo {author} {\bibfnamefont {T.}~\bibnamefont {Palstra}},
  \bibinfo {author} {\bibfnamefont {D.}~\bibnamefont {Khomskii}}, \bibinfo
  {author} {\bibfnamefont {A.}~\bibnamefont {Nugroho}}, \bibinfo {author}
  {\bibfnamefont {A.}~\bibnamefont {Menovsky}}, \ and\ \bibinfo {author}
  {\bibfnamefont {G.}~\bibnamefont {Sawatzky}},\ }\href@noop {} {\bibfield
  {journal} {\bibinfo  {journal} {Physical Review B}\ }\textbf {\bibinfo
  {volume} {62}},\ \bibinfo {pages} {6577} (\bibinfo {year}
  {2000})}\BibitemShut {NoStop}%
\bibitem [{\citenamefont {Ohkoshi}\ \emph {et~al.}(1999)\citenamefont
  {Ohkoshi}, \citenamefont {Abe}, \citenamefont {Fujishima},\ and\
  \citenamefont {Hashimoto}}]{ohkoshi99}%
  \BibitemOpen
  \bibfield  {author} {\bibinfo {author} {\bibfnamefont {S.}~\bibnamefont
  {Ohkoshi}}, \bibinfo {author} {\bibfnamefont {Y.}~\bibnamefont {Abe}},
  \bibinfo {author} {\bibfnamefont {A.}~\bibnamefont {Fujishima}}, \ and\
  \bibinfo {author} {\bibfnamefont {K.}~\bibnamefont {Hashimoto}},\ }\href@noop
  {} {\bibfield  {journal} {\bibinfo  {journal} {Physical Review Letters}\
  }\textbf {\bibinfo {volume} {82}},\ \bibinfo {pages} {1285} (\bibinfo {year}
  {1999})}\BibitemShut {NoStop}%
\bibitem [{\citenamefont {Dur\'an}\ \emph {et~al.}(2010)\citenamefont
  {Dur\'an}, \citenamefont {Ar\'evalo-L\'opez}, \citenamefont
  {Castillo-Mart\'inez}, \citenamefont {Garc\'ia-Guaderrama}, \citenamefont
  {Moran}, \citenamefont {Cruz}, \citenamefont {Fern\'andez},\ and\
  \citenamefont {Alario-Franco}}]{duran10}%
  \BibitemOpen
  \bibfield  {author} {\bibinfo {author} {\bibfnamefont {A.}~\bibnamefont
  {Dur\'an}}, \bibinfo {author} {\bibfnamefont {A.}~\bibnamefont
  {Ar\'evalo-L\'opez}}, \bibinfo {author} {\bibfnamefont {E.}~\bibnamefont
  {Castillo-Mart\'inez}}, \bibinfo {author} {\bibfnamefont {M.}~\bibnamefont
  {Garc\'ia-Guaderrama}}, \bibinfo {author} {\bibfnamefont {E.}~\bibnamefont
  {Moran}}, \bibinfo {author} {\bibfnamefont {M.}~\bibnamefont {Cruz}},
  \bibinfo {author} {\bibfnamefont {F.}~\bibnamefont {Fern\'andez}}, \ and\
  \bibinfo {author} {\bibfnamefont {M.}~\bibnamefont {Alario-Franco}},\
  }\href@noop {} {\bibfield  {journal} {\bibinfo  {journal} {Journal of Solid
  State Chemistry}\ }\textbf {\bibinfo {volume} {183}},\ \bibinfo {pages}
  {1863} (\bibinfo {year} {2010})}\BibitemShut {NoStop}%
\bibitem [{\citenamefont {Lutterotti}\ \emph {et~al.}(1999)\citenamefont
  {Lutterotti}, \citenamefont {Matthies},\ and\ \citenamefont {Wenk}}]{maud}%
  \BibitemOpen
  \bibfield  {author} {\bibinfo {author} {\bibfnamefont {L.}~\bibnamefont
  {Lutterotti}}, \bibinfo {author} {\bibfnamefont {S.}~\bibnamefont
  {Matthies}}, \ and\ \bibinfo {author} {\bibfnamefont {H.}~\bibnamefont
  {Wenk}},\ }\href@noop {} {\bibfield  {journal} {\bibinfo  {journal}
  {Newsletter of the CPD}\ }\textbf {\bibinfo {volume} {21}},\ \bibinfo {pages}
  {14} (\bibinfo {year} {1999})}\BibitemShut {NoStop}%
\bibitem [{\citenamefont {Shannon}(1976)}]{shannon76}%
  \BibitemOpen
  \bibfield  {author} {\bibinfo {author} {\bibfnamefont {R.}~\bibnamefont
  {Shannon}},\ }\href@noop {} {\bibfield  {journal} {\bibinfo  {journal} {Acta
  Crystallographica Section A}\ }\textbf {\bibinfo {volume} {32}},\ \bibinfo
  {pages} {751} (\bibinfo {year} {1976})}\BibitemShut {NoStop}%
\bibitem [{\citenamefont {Ar\'evalo-L\'opez}\ and\ \citenamefont
  {a~Alario-Franco}(2009)}]{arevalo09}%
  \BibitemOpen
  \bibfield  {author} {\bibinfo {author} {\bibfnamefont {A.}~\bibnamefont
  {Ar\'evalo-L\'opez}}\ and\ \bibinfo {author} {\bibfnamefont {M.}~\bibnamefont
  {a~Alario-Franco}},\ }\href@noop {} {\bibfield  {journal} {\bibinfo
  {journal} {Inorganic Chemistry}\ }\textbf {\bibinfo {volume} {48}},\ \bibinfo
  {pages} {11843} (\bibinfo {year} {2009})}\BibitemShut {NoStop}%
\bibitem [{\citenamefont {Dur\'an}\ \emph {et~al.}(2012)\citenamefont
  {Dur\'an}, \citenamefont {Verdin}, \citenamefont {Escamilla}, \citenamefont
  {Morales},\ and\ \citenamefont {Escudero}}]{duran12}%
  \BibitemOpen
  \bibfield  {author} {\bibinfo {author} {\bibfnamefont {A.}~\bibnamefont
  {Dur\'an}}, \bibinfo {author} {\bibfnamefont {E.}~\bibnamefont {Verdin}},
  \bibinfo {author} {\bibfnamefont {R.}~\bibnamefont {Escamilla}}, \bibinfo
  {author} {\bibfnamefont {F.}~\bibnamefont {Morales}}, \ and\ \bibinfo
  {author} {\bibfnamefont {R.}~\bibnamefont {Escudero}},\ }\href@noop {}
  {\bibfield  {journal} {\bibinfo  {journal} {Materials Chemistry and Physics}\
  }\textbf {\bibinfo {volume} {133}},\ \bibinfo {pages} {1011} (\bibinfo {year}
  {2012})}\BibitemShut {NoStop}%
\bibitem [{\citenamefont {Zhao}\ \emph {et~al.}(1993)\citenamefont {Zhao},
  \citenamefont {Weidner}, \citenamefont {Parise},\ and\ \citenamefont
  {Cox}}]{zhao93}%
  \BibitemOpen
  \bibfield  {author} {\bibinfo {author} {\bibfnamefont {Y.}~\bibnamefont
  {Zhao}}, \bibinfo {author} {\bibfnamefont {D.}~\bibnamefont {Weidner}},
  \bibinfo {author} {\bibfnamefont {J.}~\bibnamefont {Parise}}, \ and\ \bibinfo
  {author} {\bibfnamefont {D.}~\bibnamefont {Cox}},\ }\href@noop {} {\bibfield
  {journal} {\bibinfo  {journal} {Physics of the Earth and Planetary
  Interiors}\ }\textbf {\bibinfo {volume} {76}},\ \bibinfo {pages} {17}
  (\bibinfo {year} {1993})}\BibitemShut {NoStop}%
\bibitem [{\citenamefont {Sardar}\ \emph {et~al.}(2011)\citenamefont {Sardar},
  \citenamefont {Lees}, \citenamefont {Kashtiban}, \citenamefont {Sloan},\ and\
  \citenamefont {Walton}}]{sardar11}%
  \BibitemOpen
  \bibfield  {author} {\bibinfo {author} {\bibfnamefont {K.}~\bibnamefont
  {Sardar}}, \bibinfo {author} {\bibfnamefont {M.}~\bibnamefont {Lees}},
  \bibinfo {author} {\bibfnamefont {R.}~\bibnamefont {Kashtiban}}, \bibinfo
  {author} {\bibfnamefont {J.}~\bibnamefont {Sloan}}, \ and\ \bibinfo {author}
  {\bibfnamefont {R.}~\bibnamefont {Walton}},\ }\href@noop {} {\bibfield
  {journal} {\bibinfo  {journal} {Chemistry of Materials}\ }\textbf {\bibinfo
  {volume} {23}},\ \bibinfo {pages} {48} (\bibinfo {year} {2011})}\BibitemShut
  {NoStop}%
\bibitem [{\citenamefont {Alonso}\ \emph {et~al.}(2000)\citenamefont {Alonso},
  \citenamefont {Martínez-Lope}, \citenamefont {Casais},\ and\ \citenamefont
  {Fern\'andez-D\'iaz}}]{alonso00}%
  \BibitemOpen
  \bibfield  {author} {\bibinfo {author} {\bibfnamefont {J.}~\bibnamefont
  {Alonso}}, \bibinfo {author} {\bibfnamefont {M.}~\bibnamefont
  {Martínez-Lope}}, \bibinfo {author} {\bibfnamefont {M.}~\bibnamefont
  {Casais}}, \ and\ \bibinfo {author} {\bibfnamefont {M.}~\bibnamefont
  {Fern\'andez-D\'iaz}},\ }\href@noop {} {\bibfield  {journal} {\bibinfo
  {journal} {Inorganic Chemistry}\ }\textbf {\bibinfo {volume} {39}},\ \bibinfo
  {pages} {917} (\bibinfo {year} {2000})}\BibitemShut {NoStop}%
\bibitem [{\citenamefont {J\"udin}\ and\ \citenamefont
  {Sherman}(1966)}]{judin66}%
  \BibitemOpen
  \bibfield  {author} {\bibinfo {author} {\bibfnamefont {V.}~\bibnamefont
  {J\"udin}}\ and\ \bibinfo {author} {\bibfnamefont {A.}~\bibnamefont
  {Sherman}},\ }\href@noop {} {\bibfield  {journal} {\bibinfo  {journal} {Solid
  State Communications}\ }\textbf {\bibinfo {volume} {4}},\ \bibinfo {pages}
  {661} (\bibinfo {year} {1966})}\BibitemShut {NoStop}%
\bibitem [{\citenamefont {Mandal}\ \emph {et~al.}(2013)\citenamefont {Mandal},
  \citenamefont {Serrao}, \citenamefont {Suard}, \citenamefont {Caignaert},
  \citenamefont {Raveau}, \citenamefont {Sundaresan},\ and\ \citenamefont
  {Rao}}]{mandal13}%
  \BibitemOpen
  \bibfield  {author} {\bibinfo {author} {\bibfnamefont {P.}~\bibnamefont
  {Mandal}}, \bibinfo {author} {\bibfnamefont {C.}~\bibnamefont {Serrao}},
  \bibinfo {author} {\bibfnamefont {E.}~\bibnamefont {Suard}}, \bibinfo
  {author} {\bibfnamefont {V.}~\bibnamefont {Caignaert}}, \bibinfo {author}
  {\bibfnamefont {B.}~\bibnamefont {Raveau}}, \bibinfo {author} {\bibfnamefont
  {A.}~\bibnamefont {Sundaresan}}, \ and\ \bibinfo {author} {\bibfnamefont
  {C.}~\bibnamefont {Rao}},\ }\href@noop {} {\bibfield  {journal} {\bibinfo
  {journal} {Journal of Solid State Chemistry}\ }\textbf {\bibinfo {volume}
  {197}},\ \bibinfo {pages} {408} (\bibinfo {year} {2013})}\BibitemShut
  {NoStop}%
\bibitem [{\citenamefont {Yoshii}\ \emph {et~al.}(2001)\citenamefont {Yoshii},
  \citenamefont {Nakamura}, \citenamefont {Ishii},\ and\ \citenamefont
  {Morii}}]{yoshii01}%
  \BibitemOpen
  \bibfield  {author} {\bibinfo {author} {\bibfnamefont {K.}~\bibnamefont
  {Yoshii}}, \bibinfo {author} {\bibfnamefont {A.}~\bibnamefont {Nakamura}},
  \bibinfo {author} {\bibfnamefont {Y.}~\bibnamefont {Ishii}}, \ and\ \bibinfo
  {author} {\bibfnamefont {Y.}~\bibnamefont {Morii}},\ }\href@noop {}
  {\bibfield  {journal} {\bibinfo  {journal} {Journal of Solid State
  Chemistry}\ }\textbf {\bibinfo {volume} {162}},\ \bibinfo {pages} {84}
  (\bibinfo {year} {2001})}\BibitemShut {NoStop}%
\bibitem [{\citenamefont {Serrao}\ \emph {et~al.}(2005)\citenamefont {Serrao},
  \citenamefont {Kundu}, \citenamefont {Krupanidhi}, \citenamefont {Waghmare},\
  and\ \citenamefont {Rao}}]{serrao05}%
  \BibitemOpen
  \bibfield  {author} {\bibinfo {author} {\bibfnamefont {C.}~\bibnamefont
  {Serrao}}, \bibinfo {author} {\bibfnamefont {A.}~\bibnamefont {Kundu}},
  \bibinfo {author} {\bibfnamefont {S.}~\bibnamefont {Krupanidhi}}, \bibinfo
  {author} {\bibfnamefont {U.~V.}\ \bibnamefont {Waghmare}}, \ and\ \bibinfo
  {author} {\bibfnamefont {C.}~\bibnamefont {Rao}},\ }\href@noop {} {\bibfield
  {journal} {\bibinfo  {journal} {Physical Review B}\ }\textbf {\bibinfo
  {volume} {72}},\ \bibinfo {pages} {220101} (\bibinfo {year}
  {2005})}\BibitemShut {NoStop}%
\bibitem [{\citenamefont {Nogués}\ and\ \citenamefont
  {Schuller}(1999)}]{nogues99}%
  \BibitemOpen
  \bibfield  {author} {\bibinfo {author} {\bibfnamefont {J.}~\bibnamefont
  {Nogués}}\ and\ \bibinfo {author} {\bibfnamefont {I.}~\bibnamefont
  {Schuller}},\ }\href@noop {} {\bibfield  {journal} {\bibinfo  {journal}
  {Journal of Magnetism and Magnetic Materials}\ }\textbf {\bibinfo {volume}
  {192}},\ \bibinfo {pages} {203} (\bibinfo {year} {1999})}\BibitemShut
  {NoStop}%
\bibitem [{\citenamefont {Roshchin}\ \emph {et~al.}(2005)\citenamefont
  {Roshchin}, \citenamefont {Petracic}, \citenamefont {Morales}, \citenamefont
  {Li}, \citenamefont {Batlle},\ and\ \citenamefont {Schuller}}]{roshchin05}%
  \BibitemOpen
  \bibfield  {author} {\bibinfo {author} {\bibfnamefont {I.~V.}\ \bibnamefont
  {Roshchin}}, \bibinfo {author} {\bibfnamefont {O.}~\bibnamefont {Petracic}},
  \bibinfo {author} {\bibfnamefont {R.}~\bibnamefont {Morales}}, \bibinfo
  {author} {\bibfnamefont {Z.-P.}\ \bibnamefont {Li}}, \bibinfo {author}
  {\bibfnamefont {X.}~\bibnamefont {Batlle}}, \ and\ \bibinfo {author}
  {\bibfnamefont {I.}~\bibnamefont {Schuller}},\ }\href@noop {} {\bibfield
  {journal} {\bibinfo  {journal} {Europhysics Letters}\ }\textbf {\bibinfo
  {volume} {71}},\ \bibinfo {pages} {297} (\bibinfo {year} {2005})}\BibitemShut
  {NoStop}%
\bibitem [{\citenamefont {Bartolom\'e}\ \emph {et~al.}(2000)\citenamefont
  {Bartolom\'e}, \citenamefont {Bartolom\'e}, \citenamefont {Castro},\ and\
  \citenamefont {Melero}}]{bartolome00}%
  \BibitemOpen
  \bibfield  {author} {\bibinfo {author} {\bibfnamefont {F.}~\bibnamefont
  {Bartolom\'e}}, \bibinfo {author} {\bibfnamefont {J.}~\bibnamefont
  {Bartolom\'e}}, \bibinfo {author} {\bibfnamefont {M.}~\bibnamefont {Castro}},
  \ and\ \bibinfo {author} {\bibfnamefont {J.}~\bibnamefont {Melero}},\
  }\href@noop {} {\bibfield  {journal} {\bibinfo  {journal} {Physical Review
  B}\ }\textbf {\bibinfo {volume} {62}},\ \bibinfo {pages} {1058} (\bibinfo
  {year} {2000})}\BibitemShut {NoStop}%
\bibitem [{\citenamefont {Sharma}\ \emph {et~al.}(2014)\citenamefont {Sharma},
  \citenamefont {Sahoo}, \citenamefont {Perez}, \citenamefont {Mukherjee},
  \citenamefont {Gupta}, \citenamefont {Garg}, \citenamefont {Chatterjee},\
  and\ \citenamefont {Katiyar}}]{sharma14}%
  \BibitemOpen
  \bibfield  {author} {\bibinfo {author} {\bibfnamefont {Y.}~\bibnamefont
  {Sharma}}, \bibinfo {author} {\bibfnamefont {S.}~\bibnamefont {Sahoo}},
  \bibinfo {author} {\bibfnamefont {W.}~\bibnamefont {Perez}}, \bibinfo
  {author} {\bibfnamefont {S.}~\bibnamefont {Mukherjee}}, \bibinfo {author}
  {\bibfnamefont {R.}~\bibnamefont {Gupta}}, \bibinfo {author} {\bibfnamefont
  {A.}~\bibnamefont {Garg}}, \bibinfo {author} {\bibfnamefont {R.}~\bibnamefont
  {Chatterjee}}, \ and\ \bibinfo {author} {\bibfnamefont {R.}~\bibnamefont
  {Katiyar}},\ }\href@noop {} {\bibfield  {journal} {\bibinfo  {journal}
  {Journal of Applied Physics}\ }\textbf {\bibinfo {volume} {115}},\ \bibinfo
  {pages} {183907} (\bibinfo {year} {2014})}\BibitemShut {NoStop}%
\bibitem [{\citenamefont {Jacobs}\ \emph {et~al.}(1971)\citenamefont {Jacobs},
  \citenamefont {Burne},\ and\ \citenamefont {Levinson}}]{jacobs71}%
  \BibitemOpen
  \bibfield  {author} {\bibinfo {author} {\bibfnamefont {I.}~\bibnamefont
  {Jacobs}}, \bibinfo {author} {\bibfnamefont {H.}~\bibnamefont {Burne}}, \
  and\ \bibinfo {author} {\bibfnamefont {L.}~\bibnamefont {Levinson}},\
  }\href@noop {} {\bibfield  {journal} {\bibinfo  {journal} {Journal of Applied
  Physics}\ }\textbf {\bibinfo {volume} {42}},\ \bibinfo {pages} {1631}
  (\bibinfo {year} {1971})}\BibitemShut {NoStop}%
\bibitem [{\citenamefont {Yamaguchi}(1974)}]{yamaguchi74}%
  \BibitemOpen
  \bibfield  {author} {\bibinfo {author} {\bibfnamefont {T.}~\bibnamefont
  {Yamaguchi}},\ }\href@noop {} {\bibfield  {journal} {\bibinfo  {journal}
  {Journal of Physics and Chemistry of Solids}\ }\textbf {\bibinfo {volume}
  {35}},\ \bibinfo {pages} {479} (\bibinfo {year} {1974})}\BibitemShut
  {NoStop}%
\bibitem [{\citenamefont {Bertaut}(1968)}]{bertaut68}%
  \BibitemOpen
  \bibfield  {author} {\bibinfo {author} {\bibfnamefont {E.}~\bibnamefont
  {Bertaut}},\ }\href@noop {} {\bibfield  {journal} {\bibinfo  {journal} {Acta
  Crystallographica Section A: Crystal Physics, Diffraction, Theoretical and
  General Crystallography}\ }\textbf {\bibinfo {volume} {24}},\ \bibinfo
  {pages} {217} (\bibinfo {year} {1968})}\BibitemShut {NoStop}%
\bibitem [{\citenamefont {Gordon}\ \emph {et~al.}(1976)\citenamefont {Gordon},
  \citenamefont {Hornreich}, \citenamefont {Shtrikman},\ and\ \citenamefont
  {Wanklyn}}]{gordon76}%
  \BibitemOpen
  \bibfield  {author} {\bibinfo {author} {\bibfnamefont {J.}~\bibnamefont
  {Gordon}}, \bibinfo {author} {\bibfnamefont {R.}~\bibnamefont {Hornreich}},
  \bibinfo {author} {\bibfnamefont {S.}~\bibnamefont {Shtrikman}}, \ and\
  \bibinfo {author} {\bibfnamefont {B.}~\bibnamefont {Wanklyn}},\ }\href@noop
  {} {\bibfield  {journal} {\bibinfo  {journal} {Physical Review B}\ }\textbf
  {\bibinfo {volume} {13}},\ \bibinfo {pages} {3012} (\bibinfo {year}
  {1976})}\BibitemShut {NoStop}%
\bibitem [{\citenamefont {Meltzer}(1970)}]{meltzer70}%
  \BibitemOpen
  \bibfield  {author} {\bibinfo {author} {\bibfnamefont {R.}~\bibnamefont
  {Meltzer}},\ }\href@noop {} {\bibfield  {journal} {\bibinfo  {journal}
  {Physical Review B}\ }\textbf {\bibinfo {volume} {2}},\ \bibinfo {pages}
  {2398} (\bibinfo {year} {1970})}\BibitemShut {NoStop}%
\bibitem [{\citenamefont {Shamir}\ \emph {et~al.}(1981)\citenamefont {Shamir},
  \citenamefont {Shaked},\ and\ \citenamefont {Shtrikman}}]{shamir81}%
  \BibitemOpen
  \bibfield  {author} {\bibinfo {author} {\bibfnamefont {N.}~\bibnamefont
  {Shamir}}, \bibinfo {author} {\bibfnamefont {H.}~\bibnamefont {Shaked}}, \
  and\ \bibinfo {author} {\bibfnamefont {S.}~\bibnamefont {Shtrikman}},\
  }\href@noop {} {\bibfield  {journal} {\bibinfo  {journal} {Physical Review
  B}\ }\textbf {\bibinfo {volume} {24}},\ \bibinfo {pages} {6642} (\bibinfo
  {year} {1981})}\BibitemShut {NoStop}%
\bibitem [{\citenamefont {Grant}\ and\ \citenamefont {Geller}(1961)}]{grant69}%
  \BibitemOpen
  \bibfield  {author} {\bibinfo {author} {\bibfnamefont {R.}~\bibnamefont
  {Grant}}\ and\ \bibinfo {author} {\bibfnamefont {S.}~\bibnamefont {Geller}},\
  }\href@noop {} {\bibfield  {journal} {\bibinfo  {journal} {Solid State
  Communications}\ }\textbf {\bibinfo {volume} {7}},\ \bibinfo {pages} {1291}
  (\bibinfo {year} {1961})}\BibitemShut {NoStop}%
\bibitem [{\citenamefont {Kaneko}\ \emph {et~al.}(1977)\citenamefont {Kaneko},
  \citenamefont {Kurita},\ and\ \citenamefont {Tsushima}}]{kaneko77}%
  \BibitemOpen
  \bibfield  {author} {\bibinfo {author} {\bibfnamefont {M.}~\bibnamefont
  {Kaneko}}, \bibinfo {author} {\bibfnamefont {S.}~\bibnamefont {Kurita}}, \
  and\ \bibinfo {author} {\bibfnamefont {K.}~\bibnamefont {Tsushima}},\
  }\href@noop {} {\bibfield  {journal} {\bibinfo  {journal} {Journal of Physics
  C: Solid State Physics}\ }\textbf {\bibinfo {volume} {10}},\ \bibinfo {pages}
  {1979} (\bibinfo {year} {1977})}\BibitemShut {NoStop}%
\bibitem [{\citenamefont {Levison}\ \emph {et~al.}(1969)\citenamefont
  {Levison}, \citenamefont {Luban},\ and\ \citenamefont
  {Shtrikman}}]{levison69}%
  \BibitemOpen
  \bibfield  {author} {\bibinfo {author} {\bibfnamefont {S.}~\bibnamefont
  {Levison}}, \bibinfo {author} {\bibfnamefont {L.~M.}\ \bibnamefont {Luban}},
  \ and\ \bibinfo {author} {\bibfnamefont {M.}~\bibnamefont {Shtrikman}},\
  }\href@noop {} {\bibfield  {journal} {\bibinfo  {journal} {Physical Review}\
  }\textbf {\bibinfo {volume} {187}},\ \bibinfo {pages} {715} (\bibinfo {year}
  {1969})}\BibitemShut {NoStop}%
\bibitem [{\citenamefont {Yoshii}(2001)}]{yoshii01a}%
  \BibitemOpen
  \bibfield  {author} {\bibinfo {author} {\bibfnamefont {K.}~\bibnamefont
  {Yoshii}},\ }\href@noop {} {\bibfield  {journal} {\bibinfo  {journal}
  {Journal of Solid State Chemistry}\ }\textbf {\bibinfo {volume} {159}},\
  \bibinfo {pages} {204} (\bibinfo {year} {2001})}\BibitemShut {NoStop}%
\bibitem [{\citenamefont {Yoshii}(2012)}]{yoshii12}%
  \BibitemOpen
  \bibfield  {author} {\bibinfo {author} {\bibfnamefont {K.}~\bibnamefont
  {Yoshii}},\ }\href@noop {} {\bibfield  {journal} {\bibinfo  {journal}
  {Materials Research Bulletin}\ }\textbf {\bibinfo {volume} {47}},\ \bibinfo
  {pages} {3243} (\bibinfo {year} {2012})}\BibitemShut {NoStop}%
\bibitem [{\citenamefont {Cooke}\ \emph {et~al.}(1974)\citenamefont {Cooke},
  \citenamefont {Martin},\ and\ \citenamefont {Wells}}]{cooke74}%
  \BibitemOpen
  \bibfield  {author} {\bibinfo {author} {\bibfnamefont {A.}~\bibnamefont
  {Cooke}}, \bibinfo {author} {\bibfnamefont {D.}~\bibnamefont {Martin}}, \
  and\ \bibinfo {author} {\bibfnamefont {M.}~\bibnamefont {Wells}},\
  }\href@noop {} {\bibfield  {journal} {\bibinfo  {journal} {Journal of Physics
  C: Solid State Physics}\ }\textbf {\bibinfo {volume} {7}},\ \bibinfo {pages}
  {3133} (\bibinfo {year} {1974})}\BibitemShut {NoStop}%
\bibitem [{\citenamefont {Sharma}\ \emph {et~al.}(2010)\citenamefont {Sharma},
  \citenamefont {Srivastava}, \citenamefont {Krishnamurthy},\ and\
  \citenamefont {Nigam}}]{sharma10}%
  \BibitemOpen
  \bibfield  {author} {\bibinfo {author} {\bibfnamefont {N.}~\bibnamefont
  {Sharma}}, \bibinfo {author} {\bibfnamefont {B.}~\bibnamefont {Srivastava}},
  \bibinfo {author} {\bibfnamefont {A.}~\bibnamefont {Krishnamurthy}}, \ and\
  \bibinfo {author} {\bibfnamefont {A.}~\bibnamefont {Nigam}},\ }\href@noop {}
  {\bibfield  {journal} {\bibinfo  {journal} {Solid State Sciences}\ }\textbf
  {\bibinfo {volume} {12}},\ \bibinfo {pages} {1464} (\bibinfo {year}
  {2010})}\BibitemShut {NoStop}%
\bibitem [{\citenamefont {Bora}\ and\ \citenamefont {Ravi}(2015)}]{bora15}%
  \BibitemOpen
  \bibfield  {author} {\bibinfo {author} {\bibfnamefont {T.}~\bibnamefont
  {Bora}}\ and\ \bibinfo {author} {\bibfnamefont {S.}~\bibnamefont {Ravi}},\
  }\href@noop {} {\bibfield  {journal} {\bibinfo  {journal} {Journal of
  Magnetism and Magnetic Materials}\ }\textbf {\bibinfo {volume} {386}},\
  \bibinfo {pages} {85} (\bibinfo {year} {2015})}\BibitemShut {NoStop}%
\bibitem [{\citenamefont {Deng}\ \emph {et~al.}(2015)\citenamefont {Deng},
  \citenamefont {Zheng}, \citenamefont {Yu}, \citenamefont {Wang},
  \citenamefont {Sun}, \citenamefont {Avdeev}, \citenamefont {Feng},
  \citenamefont {Jing}, \citenamefont {Lu}, \citenamefont {Ren}, \citenamefont
  {Cao},\ and\ \citenamefont {Zhang}}]{deng15}%
  \BibitemOpen
  \bibfield  {author} {\bibinfo {author} {\bibfnamefont {D.}~\bibnamefont
  {Deng}}, \bibinfo {author} {\bibfnamefont {J.}~\bibnamefont {Zheng}},
  \bibinfo {author} {\bibfnamefont {D.}~\bibnamefont {Yu}}, \bibinfo {author}
  {\bibfnamefont {B.}~\bibnamefont {Wang}}, \bibinfo {author} {\bibfnamefont
  {D.}~\bibnamefont {Sun}}, \bibinfo {author} {\bibfnamefont {M.}~\bibnamefont
  {Avdeev}}, \bibinfo {author} {\bibfnamefont {Z.}~\bibnamefont {Feng}},
  \bibinfo {author} {\bibfnamefont {C.}~\bibnamefont {Jing}}, \bibinfo {author}
  {\bibfnamefont {B.}~\bibnamefont {Lu}}, \bibinfo {author} {\bibfnamefont
  {W.}~\bibnamefont {Ren}}, \bibinfo {author} {\bibfnamefont {S.}~\bibnamefont
  {Cao}}, \ and\ \bibinfo {author} {\bibfnamefont {J.}~\bibnamefont {Zhang}},\
  }\href@noop {} {\bibfield  {journal} {\bibinfo  {journal} {Applied Physics
  Letters}\ }\textbf {\bibinfo {volume} {107}},\ \bibinfo {pages} {102404}
  (\bibinfo {year} {2015})}\BibitemShut {NoStop}%
\bibitem [{\citenamefont {Zhou}\ and\ \citenamefont
  {Goodenough}(2008)}]{zhou08}%
  \BibitemOpen
  \bibfield  {author} {\bibinfo {author} {\bibfnamefont {J.}~\bibnamefont
  {Zhou}}\ and\ \bibinfo {author} {\bibfnamefont {J.}~\bibnamefont
  {Goodenough}},\ }\href@noop {} {\bibfield  {journal} {\bibinfo  {journal}
  {Physical Review B}\ }\textbf {\bibinfo {volume} {77}},\ \bibinfo {pages}
  {132104} (\bibinfo {year} {2008})}\BibitemShut {NoStop}%
\bibitem [{\citenamefont {Zhou}\ \emph {et~al.}(2010)\citenamefont {Zhou},
  \citenamefont {Alonso}, \citenamefont {Pomjakushin}, \citenamefont
  {Goodenough}, \citenamefont {Ren}, \citenamefont {Yan},\ and\ \citenamefont
  {Cheng}}]{zhou10}%
  \BibitemOpen
  \bibfield  {author} {\bibinfo {author} {\bibfnamefont {J.}~\bibnamefont
  {Zhou}}, \bibinfo {author} {\bibfnamefont {J.}~\bibnamefont {Alonso}},
  \bibinfo {author} {\bibfnamefont {V.}~\bibnamefont {Pomjakushin}}, \bibinfo
  {author} {\bibfnamefont {J.}~\bibnamefont {Goodenough}}, \bibinfo {author}
  {\bibfnamefont {Y.}~\bibnamefont {Ren}}, \bibinfo {author} {\bibfnamefont
  {J.-Q.}\ \bibnamefont {Yan}}, \ and\ \bibinfo {author} {\bibfnamefont
  {J.-G.}\ \bibnamefont {Cheng}},\ }\href@noop {} {\bibfield  {journal}
  {\bibinfo  {journal} {Physical Review B}\ }\textbf {\bibinfo {volume} {81}},\
  \bibinfo {pages} {214115} (\bibinfo {year} {2010})}\BibitemShut {NoStop}%
\bibitem [{\citenamefont {Bertaut}(1966)}]{bertaut66}%
  \BibitemOpen
  \bibfield  {author} {\bibinfo {author} {\bibfnamefont {E.}~\bibnamefont
  {Bertaut}},\ }\href@noop {} {\bibfield  {journal} {\bibinfo  {journal}
  {Journal of Applied Physics}\ }\textbf {\bibinfo {volume} {37}},\ \bibinfo
  {pages} {1038} (\bibinfo {year} {1966})}\BibitemShut {NoStop}%
\bibitem [{\citenamefont {Dong}\ \emph {et~al.}(2009)\citenamefont {Dong},
  \citenamefont {Yamauchi}, \citenamefont {Yunoki}, \citenamefont {Yu},
  \citenamefont {Liang}, \citenamefont {Moreo}, \citenamefont {Liu},
  \citenamefont {Picozzi},\ and\ \citenamefont {Dagotto}}]{dong09}%
  \BibitemOpen
  \bibfield  {author} {\bibinfo {author} {\bibfnamefont {S.}~\bibnamefont
  {Dong}}, \bibinfo {author} {\bibfnamefont {K.}~\bibnamefont {Yamauchi}},
  \bibinfo {author} {\bibfnamefont {S.}~\bibnamefont {Yunoki}}, \bibinfo
  {author} {\bibfnamefont {R.}~\bibnamefont {Yu}}, \bibinfo {author}
  {\bibfnamefont {S.}~\bibnamefont {Liang}}, \bibinfo {author} {\bibfnamefont
  {A.}~\bibnamefont {Moreo}}, \bibinfo {author} {\bibfnamefont {J.-M.}\
  \bibnamefont {Liu}}, \bibinfo {author} {\bibfnamefont {S.}~\bibnamefont
  {Picozzi}}, \ and\ \bibinfo {author} {\bibfnamefont {E.}~\bibnamefont
  {Dagotto}},\ }\href@noop {} {\bibfield  {journal} {\bibinfo  {journal}
  {Physical Review Letters}\ }\textbf {\bibinfo {volume} {103}},\ \bibinfo
  {pages} {127201} (\bibinfo {year} {2009})}\BibitemShut {NoStop}%
\end{thebibliography}%

\end{document}